\documentclass[letterpaper,11pt]{article}

\usepackage{jheppub}
\usepackage{graphicx}
\usepackage{amsmath}
\usepackage{amssymb}
\usepackage{xspace}

\newcommand{\eq}[1]{eq.~\eqref{eq:#1}}

\renewcommand{\sec}[1]{sec.~\ref{sec:#1}}

\newcommand{\subsec}[1]{sec.~\ref{subsec:#1}}
\newcommand{\app}[1]{app.~\ref{app:#1}} 
\newcommand{\fig}[1]{fig.~\ref{fig:#1}}

\newcommand{\eg}{{\it e.g.~}}
\newcommand{\ie}{{\it i.e.~}}

\newcommand{\abs}[1]{\lvert#1\rvert}

\newcommand{\ord}[1]{{\mathcal O}(#1)}
\newcommand{\ORd}[1]{{\mathcal O}\Bigl(#1\Bigr)}

\newcommand{\Mae}[3]{\bigl\langle#1\bigr\rvert#2\bigr\rvert#3\bigr\rangle}
\newcommand{\MAe}[3]{\Bigl\langle#1\Bigr\rvert#2\Bigr\rvert#3\Bigr\rangle}

\newcommand{\ket}[1]{\lvert#1\rangle}

\newcommand{\nn}{\nonumber}

\newcommand{\df}{\mathrm{d}}
\newcommand{\img}{\mathrm{i}}
\newcommand{\sdt}{\!\cdot\!}
\newcommand{\tr}{\mathrm{tr}}

\newcommand{\al}{\alpha}
\newcommand{\bt}{\beta}
\newcommand{\ga}{\gamma}
\newcommand{\Ga}{\Gamma}
\newcommand{\de}{\delta}

\newcommand{\eps}{\epsilon}
\newcommand{\ve}{\varepsilon}
\newcommand{\la}{\lambda}
\newcommand{\si}{\sigma}
\newcommand{\w}{\omega}

\newcommand{\cB}{{\mathcal B}}
\newcommand{\cG}{{\mathcal G}}
\newcommand{\cI}{{\mathcal I}}
\newcommand{\cJ}{{\mathcal J}}
\newcommand{\cL}{{\mathcal L}}
\newcommand{\cP}{{\mathcal P}}

\newcommand{\Tcm}{\mathcal{T}_\mathrm{cm}}
\newcommand{\T}{\mathcal{T}}

\newcommand{\bnslash}{\bar{n}\!\!\!\slash}

\newcommand{\ellslash}{\ell\!\!\!\slash}

\newcommand{\lp}{p_\ell}         

\newcommand{\bn}{\bar{n}}
\newcommand{\bnP}{\overline {\mathcal P}}
\newcommand{\op}{{\mathcal O}}

\newcommand{\vps}{\vec p_\perp^{\;2}}
\newcommand{\vphs}{\vec p_{h\perp}^{\;2}}
\newcommand{\vks}{\vec k_\perp^{\,2}}
\newcommand{\vtks}{\tilde {\vec k}_\perp^{\,2}}

\newcommand{\vcps}{\vec \cP_{n\perp}^{\,2}}

\newcommand{\lqcd}{\Lambda_\mathrm{QCD}}

\newcommand{\cut}{\mathrm{cut}}
\newcommand{\bare}{\mathrm{bare}}

\newcommand{\zero}{{(0)}}
\newcommand{\one}{{(1)}}

\newcommand{\SCETa}{\ensuremath{{\rm SCET}_{\rm I}}\xspace}

\newcommand{\ECM}{E_\mathrm{cm}}

\allowdisplaybreaks[2]

\title{Fully-Unintegrated Parton Distribution and Fragmentation  Functions at Perturbative $\mathbf{k_\perp}$}

\author[a]{Ambar Jain,}
\author[b]{Massimiliano Procura,}
\author[c]{Wouter J.~Waalewijn}

\affiliation[a]{Department of Physics, Carnegie Mellon University, Pittsburgh, PA~15213, U.S.A.}
\affiliation[b]{Albert Einstein Center for Fundamental Physics, Institute for Theoretical Physics, \\ University of Bern, CH-3012 Bern, Switzerland}
\affiliation[c]{Department of Physics, University of California at San Diego, 
La Jolla, CA 92093, U.S.A. \vspace{2ex}}

\emailAdd{ambar@andrew.cmu.edu}
\emailAdd{mprocura@itp.unibe.ch}
\emailAdd{wouterw@physics.ucsd.edu}

\abstract{
We define and study the properties of generalized beam functions (BFs) and fragmenting jet functions (FJFs), which are fully-unintegrated parton distribution functions (PDFs) and fragmentation functions (FFs) for perturbative $k_\perp$. 
We calculate at one loop the coefficients for matching them onto standard PDFs and FFs, correcting previous results for the BFs in the literature. Technical subtleties when measuring transverse momentum in dimensional regularization are clarified, and this enables us to renormalize in momentum space. Generalized BFs describe the distribution in the full four-momentum $k^\mu$ of a colliding parton taken out of an initial-state hadron, and therefore characterize the collinear initial-state radiation. We illustrate their importance through a factorization theorem for $pp\to \ell^+ \ell^- + 0$ jets, where the transverse momentum of the lepton pair is measured. Generalized FJFs are relevant for the analysis of semi-inclusive processes where the full momentum of a hadron, fragmenting from a jet with constrained invariant mass, is measured. Their significance is shown for the example of $e^+ e^- \to$ dijet$+h$, where the perpendicular momentum of the fragmenting hadron with respect to the thrust axis is measured.
}

\begin{document}
{\flushright INT-PUB-11-041 \\[-6ex]}

\maketitle

\section{Introduction}
\label{sec:intro}

In the description of high-energy scattering processes, the rigorous identification of the contributions from dynamics at different energy scales is achieved through factorization theorems. These allow one to systematically separate the short-distance behavior, which is calculable in perturbation theory, from process-independent, non-perturbative contributions. Furthermore, factorization enables one to sum series of large logarithms of mass scale ratios, which would otherwise make the standard perturbative expansion unreliable. 

In inclusive processes with colliding hadrons, nonperturbative effects are encoded in parton distribution functions (PDFs). The standard PDF $f_i(x,\mu)$ describes the distribution in momentum fraction $x$ of a parton of type $i=g,u, \bar u, d, \dots$ inside an incoming energetic hadron. In this paper we will discuss fully-unintegrated PDFs, which depend on all components of the four-momentum $k^\mu$ of the colliding parton. 
These allow one to keep exact kinematics 
rather than approximating 
incoming parton momenta as functions of $x$ alone.  
Consequently, realistic distributions are obtained at leading order, for detailed final state measurements, like transverse momentum and invariant mass of jets~\cite{Collins:2005uv}. 
The importance of fully-unintegrated parton densities was pointed out in the context of Monte-Carlo event generators in refs.~\cite{Watt:2003vf,Watt:2003mx,Collins:2004vq,Collins:2005uv}, where they were called ``doubly unintegrated parton densities" or ``parton correlation functions".  A field theoretic definition of fully-unintegrated PDFs and their applications in the pQCD formalism were discussed in refs.~\cite{Collins:2007ph,Rogers:2008jk}.

\begin{figure}[t]
\centering
\includegraphics[width=0.8\textwidth]{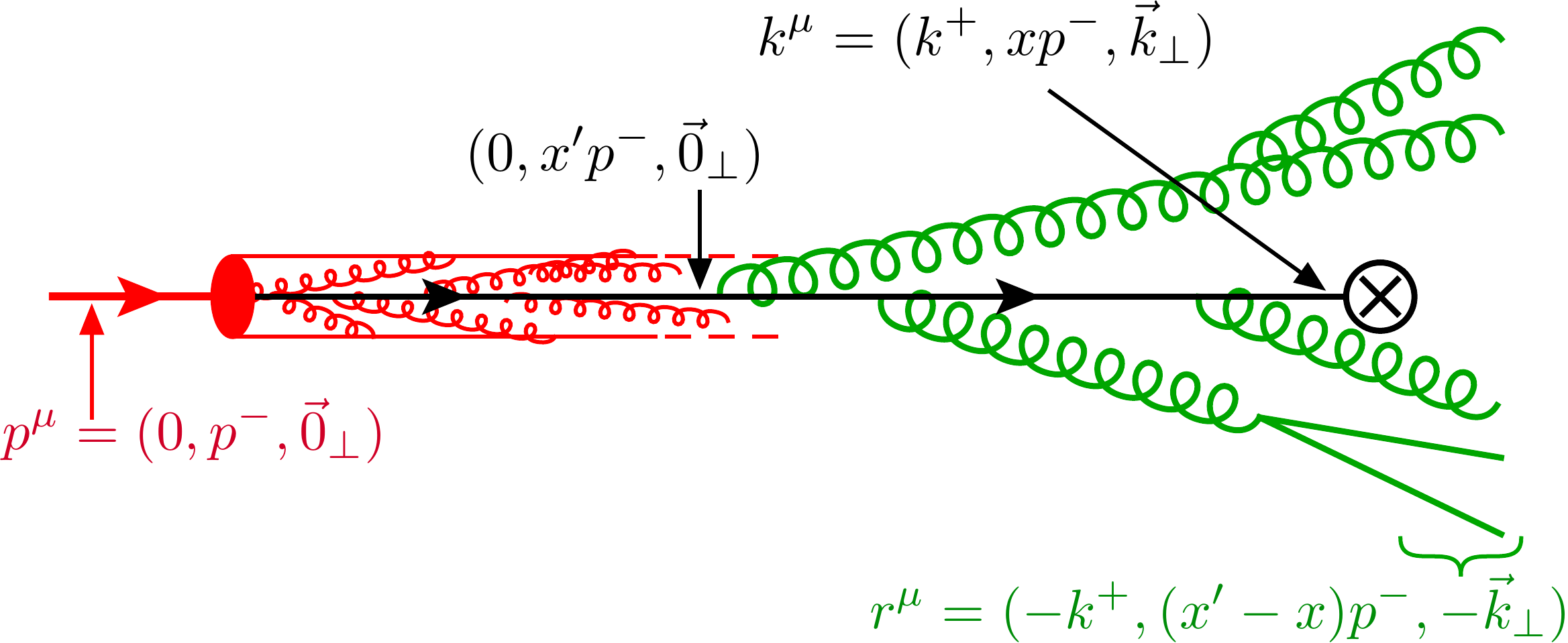}
\caption{Kinematics of the initial-state jet described by the generalized beam function.}
\label{fig:beam}
\end{figure}

Here we will focus on
hadron-hadron processes where the colliding partons are far from threshold \footnote{Here we do not resum logarithms of $1-x$, but these can be taken into account as shown in ref.~\cite{Procura:2011aq}.}
and the initial-state radiation (ISR) emitted by a parton before entering the hard subprocess is constrained to a jet along the beam axis. This can for example be imposed through a veto on central jets or by an exclusive jet measurement, see {\it e.g.}~\cite{Stewart:2009yx,Stewart:2010tn,Jouttenus:2011wh}. The momentum of the ISR is straightforwardly related to $k^\mu$, as illustrated in fig.~\ref{fig:beam} in terms of light-cone momentum components.

The generalized beam functions $B_i(t, x, \vec{k}_\perp, \mu)$ are fully-unintegrated PDFs where the parton virtuality $-t$ and transverse momentum $k_\perp$ are perturbative scales. They were defined in impact parameter space in refs.~\cite{Mantry:2009qz,Mantry:2010mk} and named iBFs. They contain information on the initial-state jets concerning both their invariant mass and their momentum component perpendicular to the beam axis, which equals $-\vec{k}_\perp$. 
The real ISR pushes the transverse virtuality $-t \equiv k^+ k^-<0$ of the colliding parton to be space-like \footnote{A proper definition of the beam functions requires a subtraction to remove the double counting of ultra-soft modes. From our private communication with the authors of refs.~\cite{Mantry:2009qz,Mantry:2010mk} we learned that they were aware of this issue, but this was not addressed in their publications.}. In our kinematic setup $\{t, \vks\} \ll Q^2$ where $Q$ is the hard scale associated with the partonic subprocess. The assumption $\{t, \vks\} \gg \lqcd^2$ allows us to calculate the dependence of the fully-unintegrated PDFs on $t$ and $\vec k_\perp$ in perturbation theory. The dependence on $x$ can be written in terms of the standard PDFs, on which the fully-unintegrated PDFs are matched. 
In this paper we will consider $t$ to be parametrically of the same size of $\vks$, which avoids dealing with the resummation of logarithms in the ratio of these two scales. 

Generalized BFs extend the definition of the standard BFs $B_i(t,x,\mu)$~\cite{Fleming:2006cd,Stewart:2009yx}, which have been extensively used to study processes with zero central jets~\cite{Stewart:2009yx, Stewart:2010pd, Berger:2010xi}, to include the measurement of the transverse recoil of ISR.
Alternatively, starting from transverse-momentum dependent (TMD) PDFs (or unintegrated PDFs)~\cite{Collins:2011ca,Cherednikov:2007tw,Hautmann:2007uw,Hautmann:2007cx,Meissner:2008xs,Pasquini:2008ax,Bacchetta:2008af,Becher:2010tm,Chiu:2012ir}, one could think of the generalized BFs as a more differential version of them. However, the generalized BFs do not suffer from the rapidity divergences that affect TMD PDFs~\cite{Collins:2003fm}, as pointed out at $\ord{\al_s}$ in ref.~\cite{Mantry:2009qz}. Instead, they have more in common with the standard BFs, like the anomalous dimension and one-loop matching onto PDFs, as we will show.
 
One of our results is that by integrating a renormalized generalized BF over $\vec k_\perp$, one recovers the corresponding renormalized standard BF. No new UV divergences appear since the integration range of $\vec k_\perp$ is bound by the measurement of $t$ and $x$:
\begin{equation} \label{eq:kin}
  t = -k^+ k^- = r^+\, x p^- = \frac{x}{1-x}\, r^+ r^- \geq \frac{x}{1-x} \vec r_{\!\perp}^{\; 2} = \frac{x}{1-x}\, \vks
\,,
\end{equation}
where $\vks \equiv -k_\perp^\mu k_{\perp\mu} \geq 0$. Here $r^\mu$ is the total momentum of the ISR (see fig.~\ref{fig:beam}), with $r^2\geq 0$ since this radiation is observed in the final state. Note that this relation between generalized and standard BFs is by definition true for bare quantities, but is nontrivial for renormalized quantities. By contrast, integrating the generalized BFs over $t$ does not yield TMD parton densities. Here the range of $t$ is not bound, so the integral over $t$ generates new divergences.

We study the renormalization of generalized BFs to all orders in perturbation theory. Their evolution is argued to be the same as for the standard beam and jet functions, which involves only the variable $t$. 
We discuss in detail how the proper definition of the relevant matrix elements is tied to the space where these get renormalized. In transverse momentum space, we show that in some cases a proper definition and renormalization requires us to measure the transverse momentum $\vec{k}_\perp$ in 2 rather than $(d-2)$ dimensions (all other momenta and polarizations are kept in $d$ dimensions). We refer to the first case as CDR$_2$ and the second case as CDR.
In the transverse coordinate space both schemes are valid and lead to the same results for the generalized BFs. These impact-parameter-dependent BFs (iBFs) were introduced in ref.~\cite{Mantry:2009qz}. 
We stress that the evolution of the BFs for a perturbative renormalization scale  does not change for nonperturbative $t$ and $k_\perp$, since the operator defining it is the same.

In this paper we present the one-loop calculation of the matching coefficients between generalized BFs and standard PDFs, according to the hierarchy $Q \gg k_\perp \simeq \sqrt{t} \gg \lqcd$:
\begin{align} \label{eq:B_OPE}
  B_i(t,x,\vks,\mu_B) 
  & = \sum_{j = u, \bar{u}, d, g \dots} \int_x^1 \frac{\df x'}{x'}
  \, \cI_{ij}\Big(t,\frac{x}{x'},\vks,\mu_B\Big) \, f_j(x',\mu_B)
   \bigg[1+ \ORd{\frac{\lqcd^2}{t}, \frac{\lqcd^2}{\vks}}\bigg]
  \,.
\end{align}
A matching of this kind was first discussed in the context of TMD fragmentation functions in ref.~\cite{Collins:1981uw}. In refs.~\cite{Fleming:2006cd,Stewart:2009yx} the analogous OPE was performed for the matching of standard beam functions onto PDFs.
In \eq{B_OPE} the matching scale $\mu_B$ should be chosen of the order of $\sqrt{t}$ to avoid large logarithms that spoil the convergence of standard perturbation theory. 
The Wilson coefficients  $\cI_{ij}$ describe the (real and virtual) radiation building up the ISR jet from the parent parton $j$ until the parton $i$ enters the hard interaction, which has the same interpretation as for the matching of standard BFs onto PDFs~\cite{Stewart:2010qs,Berger:2010xi}. We work out $\cI_{ij}$ for $i$ and $j$ being either a quark or a gluon. We will argue that the remaining ingredients (hard and soft functions) in factorization theorems for zero central jets 
are the same as those used in conjunction with standard BFs. 

Our study of matching and renormalization allows us to compare with existing results obtained for iBFs in refs.~\cite{Mantry:2009qz,Mantry:2010mk}. The diagonal coefficients $\cI_{qq}$ and $\cI_{gg}$ in \eq{B_OPE} agree with the ones obtained in ref.~\cite{Mantry:2010mk} but the mixing terms $\cI_{qg}$ and $\cI_{gq}$ do not. Taking the results of refs.~\cite{Mantry:2009qz,Mantry:2010mk}, and correcting the average over the incoming polarizations in dimensional regularization and fixing a sign resolve the discrepancy. 

In a fashion similar to beam functions, the fragmentation of a light hadron $h$ within a jet originating from a parton $i$ whose invariant mass $s$ is constrained, is described by fragmenting jet functions (FJFs) $\cG_i^h(s,z,\mu)$~\cite{Procura:2009vm}. Here $z$ denotes the hadron-parton momentum fraction.  Fragmenting jet functions have the same infrared structure as the standard fragmentation functions (FFs) $D_j^h(z,\mu)$ and can be perturbatively matched onto the latter~\cite{Jain:2011xz,Liu:2010ng}. In this framework, we have recently analyzed up to NNLL the cross section for $e^+ e^- \to X h$ on the $\Upsilon(4S)$ resonance where one restricts to the dijet limit by imposing a cut on the event shape variable thrust. This is relevant for the study of light-quark fragmentation performed by the Belle collaboration~\cite{Seidl:2008xc}. For large values of thrust we found that going beyond leading order in the cross section is important for a reliable extraction of the fragmentation function parameters~\cite{Jain:2011xz}.

Following our discussion on generalized BFs $B_i(t,x,\vec k_\perp,\mu)$, we also study the features of generalized FJFs $\cG_i^h(s,z, \vec{p}_h^{\;\perp},\mu)$, which also depend on the momentum $\vec{p}_h^{\;\perp}$ of the observed hadron perpendicular to the jet axis.  We discuss their renormalization and calculate the one-loop matching coefficients onto standard FFs according to the hierarchy $Q \gg p_h^{\;\perp} \simeq \sqrt{s} \gg \lqcd$. We argue that to all orders in perturbation theory the running of generalized FJFs is the same as that of the jet functions (as is the case for the standard FJFs).

The paper is organized as follows. In \sec{renorm} we set up the theoretical framework, give the definitions of quark and gluon generalized BFs and FJFs, and discuss their all-order renormalization, both in momentum space and in impact parameter space. We make a general comparison with the existing SCET literature on generalized BFs~\cite{Mantry:2009qz} in \subsec{compiBFs}, and discuss the disagreement on the matching coefficients in \sec{sonny}. In \app{qfupdf} we stress the need for ultra-soft zero-bin subtractions in the definition of BFs via a detailed calculation of the quark generalized BF where UV, IR and rapidity divergences are taken care of by different regulators. Sec.~\ref{sec:oneloop} is devoted to the calculation of the quark and gluon matching coefficients onto the standard PDFs/FFs. The details of the one-loop calculation of the gluon BF in momentum space using CDR$_2$ are given in \app{gfupdf}. In \sec{factth}  (and \subsec{ren_BF}) we illustrate the relevance of generalized BFs and FJFs with factorization theorems for transverse momentum dependent distributions. We discuss $p_\perp$-distributions in Drell-Yan with a central jet veto and in $gg \to H \to W^+ W^- \to \ell^+ \nu \ell'^- \bar{\nu}'+ 0\; {\rm jets}$. We also analyse single-hadron fragmentation in $e^+ e^-$ where a cut on thrust is used to restrict to dijet final states and the full hadron momentum is measured.

\section{Definition and Renormalization of Generalized BFs and FJFs}
\label{sec:renorm}

\subsection{SCET Ingredients}
\label{subsec:SCET}

As explained in the introduction, in this paper we deal with processes governed by three different perturbative scales: a hard scale $Q$ associated with the partonic subprocess, an intermediate jet scale given by the square root of the jet invariant mass $s$ or the transverse virtuality $t$, and a soft scale of order  $s/Q$  or $ t/Q$, respectively. We will always consider the mass of the incoming/fragmenting hadron to be negligible. We will also include effects at the non-perturbative scale $\lqcd$ through standard PDFs and FFs.
We restrict our analysis to the case that the transverse momentum of the ISR or the momentum of the fragmenting hadron perpendicular to the jet axis are of the order of the jet scale. After integrating out the hard dynamics, we are left with degrees of freedom describing the energetic radiation inside well-separated jets and the soft emission between them (collinear and ultra-soft modes, respectively). Therefore, Soft-Collinear Effective Theory (SCET)~\cite{Bauer:2000ew, Bauer:2000yr, Bauer:2001ct, Bauer:2001yt} is well suited to study generalized BFs and FJFs. 

Collinear modes are characterized by having large energy and small invariant mass, and therefore are conveniently described using light-cone coordinates. 
We introduce a light-cone vector $n^\mu$ whose spatial part is along the collimation axis, and another light-cone vector $\bn^\mu$ such that $n^2 = \bn^2 = 0$ and $n\cdot\bn = 2$. 
Any four-vector $p^\mu$ can then be decomposed as $p^\mu = (p^+, p^-, p^\mu_\perp)$ with $p^+ = n\cdot p$, $p^- = \bn\cdot p$ and $p_\perp^\mu$, which contains the components of $p^\mu$ perpendicular to $n^\mu$ and $\bn^\mu$. The momentum $p^\mu$ of a collinear particle scales as $(p^+, p^-, p_\perp^\mu) \sim p^- (\la^2, 1, \la)$, where $\la \ll 1$ is the SCET expansion parameter. For the ultra-soft degrees of freedom, the momentum scales like $q^\mu = (q^+, q^-, q_\perp^\mu) \sim p^- (\la^2, \la^2, \la^2)$. 

The SCET fields for $n$-collinear quarks and gluons, $\xi_{n,\lp}(y)$ and $A_{n,\lp}(y)$ respectively, are labeled by the direction $n$ and the large (discrete) momentum $\lp$. Their argument $y$ is conjugate to the small residual (continuous) momenta \footnote{Throughout this paper, for notational convenience, we combine residual $p_r$ and label $\lp$ momenta into a continuous variable $p$: for example, $\de_{\w,\lp^-}\,\de(p_r^-) \equiv \de(\w - p^-)$ and similarly for the perpendicular components.}.
We explicitly exclude the case $\lp^\mu = 0$ in collinear fields to avoid double-counting the soft degrees of freedom (which are described by separate ultra-soft quark and gluon fields). In practice, when calculating matrix elements, this is implemented via zero-bin subtractions~\cite{Manohar:2006nz}. 

Collinear operators are built out of products of fields and Wilson lines that are invariant under collinear gauge transformations~\cite{Bauer:2000yr,Bauer:2001ct}. The basic building blocks are the collinearly gauge-invariant quark and gluon fields, defined as
\begin{equation} \label{eq:chiB}
\chi_n(y) = W_n^\dagger(y)\, \xi_n(y)
\,,\qquad
\cB_{n\perp}^\mu(y) = \frac{1}{g} \bigl[W_n^\dagger(y)\, \img D_{n\perp}^\mu W_n(y) \bigr]
\,,\end{equation}
where $\img D_{n\perp}^\mu = \cP^\mu_{n\perp} + g A^\mu_{n\perp}$ is the $\perp$-collinear covariant derivative. The collinear Wilson line
\begin{equation} \label{eq:Wn}
W_n(y) = \biggl[\sum_\text{perms} \exp\Bigl(-\frac{g}{\bnP_n}\,\bn\sdt A_n(y)\Bigr)\biggr]
\end{equation}
sums up arbitrary emissions of $n$-collinear gluons which are $\ord{1}$ in the power counting.

At leading order in the SCET power expansion, the interactions of ultra-soft gluons with collinear fields exponentiate to form eikonal Wilson lines. The ultra-soft gluons can thus be decoupled via the BPS field redefinition~\cite{Bauer:2001yt}
\begin{align} \label{eq:BPS}
\chi^\zero_n(y) &= Y_n^\dagger(y)\,\chi_n(y)
\,, \nn \\
\cB^{\mu\zero}_{n\perp}(y) &= Y_n^\dagger(y)\,\cB^\mu_{n\perp}(y)\,Y_n(y)
\,.\end{align}
The collinear fields we consider in this paper are those after this decoupling, and we drop the superscript $(0)$ for notational convenience. Here $Y_n(y)$ is an ultra-soft Wilson line in the fundamental representation 
\begin{align} \label{eq:Yn}
Y_n(y) &=  P\exp\biggl[\,\img g\int_{-\infty}^0 \df u\, n\sdt A_{us}(y + u\,n) \biggr]  
\,.
\end{align}
The symbol $P$ in \eq{Yn} denotes path ordering of the color generators along the integration path.

\subsection{Generalized Beam Functions}
\label{subsec:GBF}

\subsubsection{Definition}

The bare quark and gluon beam functions (BFs) are defined by the following proton matrix elements of bare SCET operators:
\begin{align} \label{eq:B_def}
B^\bare_q&(t,x,\vks)
= \\
& \theta(k^-) \, 
\Mae{p_n(p^-)}{\bar\chi_n(0)\, \de(t - k^- \hat p^+) \frac{\bnslash}{2} 
  \big[\delta(k^- -\bnP_n)\, \frac{1}{\pi} \de(\vks - \vcps)  \, \chi_n(0)\big]\,}{p_n(p^-)}
\, , \nn \\
B^{\mu \nu,\,\bare}_g&(t,x,\vec{k}_\perp)
= 
\nn \\ & 
- k^- \, \theta(k^-) 
\Mae{p_n(p^-)}{\cB_{n\perp}^{\mu c}(0)\, \de(t - k^- \hat p^+) 
  \big[\delta(k^- -\bnP_n) \de^2(\vec k_\perp - \vec \cP_{n\perp})\, \cB_{n\perp}^{\nu c}(0) \big]}{p_n(p^-)}
\,, \nn
\end{align}
where an average over the proton spin is assumed. 
The light-like vector $n^\mu$ is chosen such that the proton states have no perpendicular momentum, $p^\mu = p^- n^\mu /2$. By boost invariance along the $z$-axis, these functions only depend on the momentum fraction $x= k^-/p^-$, the transverse virtuality $-t =  k^- k^+$ of the parton, and the transverse momentum $\vec{k}_\perp$. 
The label momentum operators $\cP$ act on the fields inside the square brackets whereas $\hat p^+$ measures the plus momentum of any intermediate state.

If we switch from momentum to position space, the fields in \eq{B_def} will be separated by a distance $y^\mu$. Ultra-soft Wilson lines along the large $y^-$ separation are factored out in the BPS-redefined fields $\chi_n$ and $\cB_{n\perp}$ of \eq{B_def}. The collinear Wilson lines in \eq{chiB} are along the small $y^+$ direction. Since $y_\perp \neq 0$, the BFs defined in \eq{B_def} are invariant under (ultra-soft and collinear) covariant gauge transformations because in this case the gauge field vanishes at infinity.
The subtleties arising in the definition of these functions in singular gauges are similar to those affecting TMD PDFs, see \eg refs.~\cite{Ji:2002aa, Belitsky:2002sm, Collins:2003fm, Idilbi:2008vm, Idilbi:2010im, GarciaEchevarria:2011md}. Since we deal with factorization theorems for gauge invariant quantities, we can always choose a non-singular gauge to carry out our calculations.

The quark BF is a Lorentz scalar which only depends on the magnitude of $\vec{k}_\perp$ and hence we could replace $\de^2(\vec k_\perp - \vec \cP_{n\perp})$ with $\de(\vks - \vcps)/\pi$. In the gluon BF, the measurement of $\vec k_\perp$ allows a new Lorentz structure~\cite{Mantry:2009qz}, namely
\begin{align} \label{eq:Lor}
&B_g^{\mu\nu}(t,x,\vec{k}_\perp) =  B_1(t,x,\vks)\, L_1^{\mu\nu} + B_2(t,x,\vks)\, L_2^{\mu\nu}(\vec k_\perp)
\,,\nn \\ 
&L_1^{\mu \nu} = \frac{g_\perp^{\mu \nu}}{2}\,, \qquad
 L_2^{\mu \nu}(\vec k_\perp) = \frac{k_\perp^\mu k_\perp^\nu}{\vks} + \frac{g_\perp^{\mu \nu}}{2}\,, \qquad
 L_1^{\mu \nu}  L_{2\mu \nu}(\vec k_\perp) = 0
 \,.
\end{align}
Thus the tensor structure of the gluon BF and the dependence on the full $\vec{k}_\perp$ must be kept. In a factorized cross section for two colliding gluons the Lorentz indices between two gluon BFs get contracted, see \eq{factren}. The new Lorentz structure $L_2$ only starts at one loop in the BF and it only starts to contribute to the cross section at two loops, since $L_1 \cdot L_2 = 0$.

In order to interpret \eq{B_def} for the quark BF in QCD, in the Fourier transform, the field $\chi_n(y)$ should be replaced by $V(y,n) \psi(y)$ with $V(y,n)$ defined as $P \exp \big(-i g \int_0^\infty ds \,  n \cdot\! A(y + s n) \big)$.
In addition, soft subtractions need to be performed to avoid double counting of overlapping momentum regions~\cite{Collins:1999dz}, in analogy to SCET zero-bin subtractions. 

An alternate definition of fully unintegrated parton distribution function, dubbed parton correlation function (PCF), was given in ref.~\cite{Collins:2007ph} in the context of QCD factorization. In covariant gauges and in coordinate space, the PCF is defined as
\begin{align}\label{eq:pennstatePCF}
\tilde F(y, \eta_p, \eta_s) =    \frac{\langle p \vert \bar \psi(y) V^\dagger(y,n_s) \frac{\bnslash}{2}  V(0,n_s) \psi(0) \vert p \rangle}{\langle 0 \vert V(y, n_T)V^\dagger(y,n_s)V(0, n_s)V^\dagger(0,n_T) \vert 0\rangle}
\end{align}
which depends on all four components of the coordinate $y$, proton rapidity $\eta_p$ and a soft ``cut-off'' rapidity $\eta_s$. $n_s = (-e^{\eta_s},e^{-\eta_s},{\mathbf 0}_\perp)$ is a space-like vector required to provide a boundary between left-moving and right-moving partons. $n_T = (- e^{-\vert 2 \eta_p \vert},1, {\mathbf 0}_\perp)$ is a non-light-like (or approximately light-like) vector and is required to regulate the rapidity divergences. Qualitatively, the denominator serves the same purpose as zero-bin subtractions in SCET. However, there are important distinctions between our definition and the PCF. Note that the numerator in eqn. (\ref{eq:pennstatePCF}), which can be considered as a naive definition of the PCF, has space-like Wilson lines unlike our definition of BFs which has light-like Wilson lines. Another important distinction arises in the RG equations. Apart from the usual RG equation in $\mu$, the PCF satisfies a Collins-Soper equation in $\eta_s$. This $\eta_s$ dependence would cancel against another collinear sector, so that the cross-section does not depend on $\eta_s$. In our case, the soft overlap of the beam functions is only with momenta of order $Q(\lambda^2,\lambda^2,\lambda^2)$, hence BFs are not rapidity divergent after zero-bin subtractions have been implemented. This is discussed in Appendix \ref{app:qfupdf}. As a consequence, the standard RG equation in $\mu$ is sufficient to sum all the large logarithms in our BFs.

\subsubsection{Treating Transverse Momenta in Dimensional Regularization}
\label{subsec:HVvsCDR}

In this section we discuss a technical issue arising in dimensional regularization when the full perpendicular momentum is measured rather than its norm. This is the case, for example, of the gluon generalized BF. We now show that the proper definition of this function {\it in momentum space} involves the measurement of $\vec{k}_\perp$ in 2 dimensions (CDR${}_2$) rather than in $d-2$ dimensions (CDR). With renormalization carried out in impact parameter space, CDR and CDR${}_2$ are both valid schemes which yield the same results for the calculations in this paper, as shown in \sec{sonny}. However, we expect that CDR and CDR${}_2$ would in general lead to different results for the finite terms of other transverse-coordinate-dependent functions.

The essential point of our argument is that the matrix element of operators (like the BFs) should be in integer dimensions, to have an unambiguous expansion in $\eps = (4-d)/2$. This cleanly separates the divergences, facilitating the resummation of logarithms sector by sector. Requiring the matrix elements to be in integer dimensions fixes the dimension of the $\de$-function measuring the transverse momentum.

We point out that the bare matrix elements have the same integer mass dimension as the renormalized ones. This follows from
\begin{align}\label{eq:op-ren}
\langle{\cal O}_{\rm ren}\rangle & = Z_{\cal O}^{-1} \otimes \langle {\cal O}_{\rm bare}\rangle = (Z_{\phi_1}^{1/2}\, Z_{\phi_2}^{1/2} \cdots ) \,  Z_{\cal O}^{-1} \otimes \langle \phi_1\phi_2\cdots\rangle \, ,
\end{align} 
where $Z_{\phi_i}$ relates bare and renormalized ($\phi_i$) fields and $Z_{\cal O}$ is the operator renormalization factor. 
The mass dimensions of the $Z$'s and of the integration measure in the convolution ``$\otimes$" cancel each other, as is clear at tree level.

Let us consider the example of \eq{B_def}. For the fields, $[\chi_n] = 3/2-\epsilon$ and $[\cB_{n\perp}] = 1-\eps$. For a single particle in $d$ dimensions, $ [ \ket{p} ] =-1+\eps$. 
In \eq{B_def} the $\eps$-dimensions thus cancel between fields and external one-particle states. Therefore the transverse momentum for the generalized gluon BF has to be measured in 2 dimensions, thus CDR${}_2$ is a sensible scheme. We stress that this would not be the case in CDR, where \eq{B_def} would involve a $\de^{d-2}(\vec k_\perp - \vec \cP_{n\perp})$. 
For the quark generalized BF, which depends on the norm $|\vec{k}_\perp|$, both schemes will work. In CDR this requires using the identity 
\begin{align}
\delta(\vec k_\perp^2-\vec{\cal P}_{n \perp}^2)/\pi = \frac{(\vec k_\perp^2)^{-\epsilon}}{\Gamma(1-\epsilon)\pi^\epsilon} \,\delta^{2-2\epsilon}(\vec k_\perp - \vec{\cal P}_{n\perp})
\end{align}

In impact parameter space the gluon generalized BF is renormalizable both in CDR${}_2$ and in CDR. In the latter case one performs a $(d-2)$-dimensional Fourier transform with respect to $\vec k_\perp$~\cite{Mantry:2009qz,Mantry:2010mk} which ensures that the bare iBF is in integer dimensions. 	
The proper modifications of the Lorentz structures in \eq{Lor} for the bare beam functions in the CDR and CDR${}_2$ schemes are 
\begin{align} \label{eq:Lor2}
\text{CDR}: & \qquad 
L_1^{\mu \nu} = \frac{g_\perp^{\mu \nu}}{d-2}\,, \qquad
 L_2^{\mu \nu}(\vec k_\perp) = \frac{k_\perp^\mu k_\perp^\nu}{\vks} + \frac{g_\perp^{\mu \nu}}{d-2}
 \,,\nn \\
\text{CDR}{}_2: & \qquad 
L_1^{\mu \nu} = \frac{g_\perp^{\mu \nu}}{d-2}\,, \qquad
 L_2^{\mu \nu}(\vec k_\perp) = \frac{k_\perp^\mu k_\perp^\nu}{\vks} + \frac{g_2^{\mu \nu}}{2}
 \,,
\end{align}
where $k_\perp^\mu$  is $(d-2)$ dimensional in CDR and two dimensional in CDR${}_2$. Here $g_2^{\mu \nu}$ denotes the purely two-dimensional piece of $g_\perp^{\mu\nu}$. Since in CDR${}_2$ only the purely two-dimensional perpendicular momentum gets measured, the $-2\eps$-dimensional contribution is not associated to a specific direction and should only appear in $L_1$ . This is shown in the explicit calculation for the gluon BF in \app{gfupdf}.

\subsubsection{General Comparison with SCET Literature}
\label{subsec:compiBFs}

Generalized BFs are purely collinear matrix elements composed of collinear fields and states. Integrations over full phase space and loop momenta overlap with the momentum region corresponding to the ultra-soft modes. To attain a proper definition of collinear matrix elements we must implement zero-bin subtractions~\cite{Manohar:2006nz}. Previously, generalized BFs in impact parameter space (iBFs) were defined without explicit ultra-soft zero-bin subtractions~\cite{Mantry:2009qz,Mantry:2010mk} \footnote{In our private communications the authors of these references agree that the proper definition of the iBF requires ultra-soft zero-bin subtractions.}. We show in \app{qfupdf} that in absence of these subtractions the generalized BFs suffer from rapidity divergences, by explicitly calculating the quark BF at one-loop with IR and rapidity regulators different from dimensional regularization. 

Generalized BFs naturally arise in the context of observables which are sensitive to collinear and ultra-soft modes. In our framework the initial-state radiation is constrained to energetic jets (described by collinear fields) with only ultra-soft radiation between jets due to our measurement of the transverse virtuality $-t \sim Q^2 \lambda^2$. Therefore the soft modes [which scale like $Q(\lambda, \lambda, \lambda)$] cannot appear here as real radiation. Virtual exchange of soft gluons could in principle transfer perpendicular momentum between the two colliding partons. However, there is no measurement sensitive to this momentum transfer, rendering scaleless integrals for the soft contribution. In other words, a collinear particle within a jet can only recoil against another particle within the same jet. Our observables will become sensitive to soft radiation only if parametrically $\vks \ll t \sim \lambda Q^2$. However, in this case the contribution of collinear radiation to $t$ is power suppressed and one would have to deal with TMD PDFs rather than generalized beam functions. Alternatively, the generalized BFs do not have support on the soft region. We have assumed that effects from Glauber modes cancel out.

By contrast in refs.~\cite{Mantry:2009qz,Mantry:2010mk}, the iBFs were used for cross sections where there is no constraint on $t$ but only on $k_\perp$. This in principle permits contributions from soft modes, which they treat as explicit degrees of freedom in the effective theory. In light of the previous paragraph this raises concern about the proper accounting of modes and power counting in their factorization theorem. 
For example, their soft function depends on all momentum components, so it naively has an overlap with the ultra-soft region. Since the ultra-soft zero-bin subtractions were performed for the iBFs, which also depend on all momentum components, it is not clear to us how the ultra-soft subtractions for the soft function should be performed in their framework. A detailed discussion of their factorization theorem is beyond the scope of this paper.

In \sec{sonny} we will resolve a discrepancy in the literature between refs.~\cite{Mantry:2010mk,Berger:2010xi} regarding the matching coefficients $\cI_{qg}$ and $\cI_{gq}$ of BFs onto PDFs, in favor of ref.~\cite{Berger:2010xi}. In pure dimensional regularization zero-bins vanish and therefore the matching coefficients obtained in ref.~\cite{Mantry:2009qz,Mantry:2010mk} should coincide with our calculation in this paper. As we will explain, the discrepancy is due to an oversight in ref.~\cite{Mantry:2010mk}.

\subsubsection{Renormalization}
\label{subsec:ren_BF}

We will now argue that the renormalization of the generalized BF is the same as that of the standard BF, to all orders in perturbation theory. In turn this equals the jet function renormalization~\cite{Stewart:2010qs} for which the anomalous dimension is known to three-loop order~\cite{Becher:2006mr,Becher:2009th,Berger:2010xi}. This fact is quite useful since resummed calculations require anomalous dimensions at higher order in $\al_s$ than the fixed-order contribution. 
 
Assuming the validity of a factorization theorem, we give an argument for the all-orders renormalization of the generalized BFs. 
In essence, in the factorization theorem the additional transverse momentum dependence only appears in the BF, and thus the generalized and standard BF have the same anomalous dimension. Consider, for example, the case of Higgs production through gluon fusion ($gg \to H$), where a central jet veto is imposed through the beam thrust event shape $\Tcm$ defined in the hadronic center-of-mass frame~\cite{Berger:2010xi}. This example involves the two different Lorentz structures in the gluon BF. It will be convenient to separately measure the contributions to $\Tcm$ from each of the two hemispheres orthogonal to the beam axis defined as
\begin{equation}
 \Tcm = \T_a+\T_b
 \,, \qquad
\T_a
= \sum_i\, \theta(\eta_i) \abs{\vec p_i^{\,\perp}}\, e^{-\eta_i}
 \,, \qquad
\T_b
= \sum_i\, \theta(-\eta_i) \abs{\vec p_i^{\,\perp}}\, e^{\eta_i}
\,,\end{equation}
where the sum on $i$ runs over all particles in the final state except the Higgs. Here  $|\vec p_i^{\,\perp}|$ and $\eta_i$ are the transverse momentum and pseudorapidity of the particle $i$ with respect to the beam axis.
The factorization theorem for small beam thrust in ref.~\cite{Stewart:2009yx} can be generalized to the case where we also measure the transverse momenta of the two ISR jets $\vec k_a^\perp = \sum_i \theta(\eta_i) \vec p_i^\perp$, $\vec k_b^\perp = \sum_i \theta(-\eta_i) \vec p_i^\perp$,
\begin{align} \label{eq:factren}
&\frac{\df\sigma}{\df \T_a\, \df \T_b\, \df \vec k_a^\perp \, \df  \vec k_b^\perp\, \df Y}
\\ & \quad
= \sigma_0\, H_{gg}(m_H^2, \mu) \int\!\df t_a\, \df t_b\, 
S_\text{ihemi}^{gg}\Bigl(\T_a \!-\! \frac{e^{-Y} t_a}{m_H}, \T_b \!-\! \frac{e^Y t_b}{m_H}, \mu\Bigr)
B_g^{\mu\nu}(t_a, x_a, \vec{k}_a^\perp, \mu)\, B_{g\,\mu\nu}(t_b, x_b, \vec{k}_b^\perp,\mu)
\nn \\ & \quad
= \sigma_0\, H_{gg}(m_H^2, \mu) \int\!\df t_a\, \df t_b\, 
S_\text{ihemi}^{gg}\Bigl(\T_a \!-\! \frac{e^{-Y} t_a}{m_H}, \T_b - \frac{e^Y t_b}{m_H}, \mu\Bigr)
\nn\\ &\quad \ \times\!
\bigg\{\frac{1}{2} B_1(t_a,x_a, \vec k_{a\perp}^{\,2},\mu) B_1(t_b,x_b,\vec k_{b\perp}^{\,2},\mu) \!+\! \bigg[\frac{(\vec k_a^\perp \sdt \vec k_b^\perp)^2}{\vec k_{a\perp}^{\,2} \vec k_{b\perp}^{\,2}} \!-\! \frac{1}{2}\bigg] B_2(t_a,x_a, \vec k_{a\perp}^{\,2},\mu) B_2(t_b,x_b, \vec k_{b\perp}^{\,2},\mu) \bigg\}
,\nn\end{align}
where $\si_0$ is the Born cross section.
The rapidity $Y$ and mass $m_H$ of the Higgs are related to the momentum fractions $x_{a,b}$ by
\begin{equation}
x_a = \frac{m_H}{\ECM}\,e^{Y}
\,,\qquad
x_b = \frac{m_H}{\ECM}\,e^{-Y}
\,.
\end{equation}
The second equality in \eq{factren} is obtained by inserting the Lorentz structures of \eq{Lor}.
The hard function $H_{gg}$ describes the virtual corrections at the hard scale $m_H$, and $S_\text{ihemi}^{gg}$ is the incoming hemisphere soft function.
Since only the large momentum components enter the hard function $H_{gg}$, the additional measurement of the transverse momenta $\vec k_{a\perp}$, $\vec k_{b\perp}$ does not affect it. By requiring that parametrically $\{\T_a,\T_b\} \sim \{\vec k_{a\perp},\vec k_{b\perp}\}$, the contribution of the ultra-soft radiation to the transverse momenta is power suppressed, so  $S_\text{ihemi}^{gg}$  is the same soft function as in ref.~\cite{Stewart:2009yx}. Soft degrees of freedom do not contribute, as explained in \subsec{compiBFs}. 

The $\mu$-dependence of the cross section cancels up to the order one is working, implying that the anomalous dimensions of the hard, soft and beam functions cancel each other. Since $H_{gg}$ and $S_\text{ihemi}^{gg}$ are unchanged by the additional measurement of the transverse momenta, the products of the standard beam functions $B_g(t_a,x_a,\mu) B_g(t_b,x_b,\mu)$ and the products of the generalized beam functions $B_1(t_a,x_a, \vec k_a^{\,2},\mu) B_1(t_b,x_b, \vec k_b^{\,2},\mu)$, $B_2(t_a,x_a, \vec k_a^{\,2},\mu) B_2(t_b,x_b, \vec k_b^{\,2},\mu)$ all have the same running in $\mu$. The relative contribution of the $B_1 B_1$ and $B_2 B_2$ Lorentz structures can be varied through the angle between $\vec k_a$ and $\vec k_b$, so they are independent and cannot mix under renormalization. Since the variables $t_a,t_b,x_a,x_b,\vec k_a^{\,2}, \vec k_b^{\,2}$ can be all independently varied through $\T_a, \T_b, m_H, Y, \vec k_a^{\,2}, \vec k_b^{\,2}$, it follows that the generalized gluon BF has the same renormalization as the standard gluon BF. The same is true for the quark beam function. 

The renormalization of the standard beam function equals that of the jet function~\cite{Stewart:2010qs}, so we conclude that the renormalization of the generalized beam functions is given by
\begin{align}  \label{eq:B_ren}
  B_i^\bare (t,x, \vec k_\perp) & = \int_0^t\! \df t'\, Z_B^i(t - t',\mu)\, B_i(t',x, \vec k_\perp,\mu)
  \,, \qquad
  Z_B^i(t,\mu) =   Z_J^i(t,\mu)
  \,, \nn \\
  \mu \frac{\df}{\df \mu} B_i(t, x, \vec k_\perp, \mu) & = \int_0^t\! \df t'\, \ga_B^i(t-t',\mu)\,  B_i(t', x, \vec k_\perp, \mu)
  \,, \qquad
  \ga_B^i(t,\mu) =  \ga_J^i(t,\mu)
  \,,
\end{align}
where $Z_J$ and $\ga_J$ are the jet function renormalization factor and anomalous dimension. There is no mixing between different parton types, so $i$ on the right-hand side is fixed (not summed). Explicit expressions for $\ga_B^q$ may be found in app.~D.2 of ref.~\cite{Stewart:2010tn} and for $\ga_B^g$ in app.~B.3 of ref.~\cite{Berger:2010xi}.

It may be surprising that the renormalization in \eq{B_ren} depends on $t = -k^- k^+$, instead of the Lorentz invariant combination $t + \vks=-k^2$. In SCET, Lorentz invariance is broken by the choice of $n^\mu$ and $\bn^\mu$. Instead we have reparametrization invariance (RPI) \cite{Chay:2002vy, Manohar:2002fd}, which encodes the arbitrariness in choosing $n^\mu$ and $\bn^\mu$. 
RPI can be divided into three types: RPI-I and RPI-II transformations correspond to rotations of $n$ and $\bn$, while for RPI-III
$n^\mu \to e^\alpha\, n^\mu\, , \quad \bn^\mu \to e^{-\alpha}\, \bn^\mu\, $.
These transformations preserve the SCET power counting and the defining relations $n^2 = \bn^2 =0$, $n \cdot \bn = 2$.
Let us now study the implications of RPI on the renormalization of the quark beam function. To this end we strip the beam function in \eq{B_def} of the external proton states, since the states do not affect the renormalization. Starting from the most general structure for the renormalization of the resulting operator $\op_q$, we find
\begin{align} \label{eq:Z_RPI}
 \op_q^\bare(k^-,k^+,\vks) &= \int\! \df \tilde k^- \df \tilde k^+ \df  \vtks\,
 Z_q(k^-,k^+,\vks,\tilde k^-,\tilde k^+,\vtks,\mu)\, \op_q(\tilde k^-,\tilde k^+, \vtks,\mu) \nn \\
&= \int\! \df \tilde k^- \df \tilde k^+ \df \vtks \, 
 \bigg[\de\Big(1-\frac{k^-}{\tilde k^-}\Big) \de(\vks - \vtks) Z_1^q(k^- k^+,\tilde k^- \tilde k^+,\mu)
 \nn \\ & \hspace{19ex}
 +  Z_2^q\Big(\frac{k^-}{\tilde k^-}, k^2 , \tilde k^2,\mu\Big) \bigg]\op_q(\tilde k^-,\tilde k^+,\vtks,\mu)\,.
\end{align}
The RPI transformations change $\bnP$, $\cP_{n\perp}^\mu$ and $\hat p^+$ inside $\op_q$ into one other. By a suitable change of variables for $k^-,k^+, \vks$ and $\tilde k^- ,\tilde k^+, \vtks$, the operators take their old form and the effect of the RPI transformation is moved entirely into the $Z$ factors. Based on RPI two structures are allowed: in $Z_1^q$ the renormalization is purely in $t$, in $Z_2^q$ the renormalization can depend on both $k^2$ and $z$. Eq.~\eqref{eq:B_ren} shows that the structure $Z_2^q$ is absent to all orders, as verified by our one-loop calculation.

\subsection{Generalized Fragmenting Jet Functions}
\begin{figure}[t]
\centering
\includegraphics[width=0.6 \textwidth]{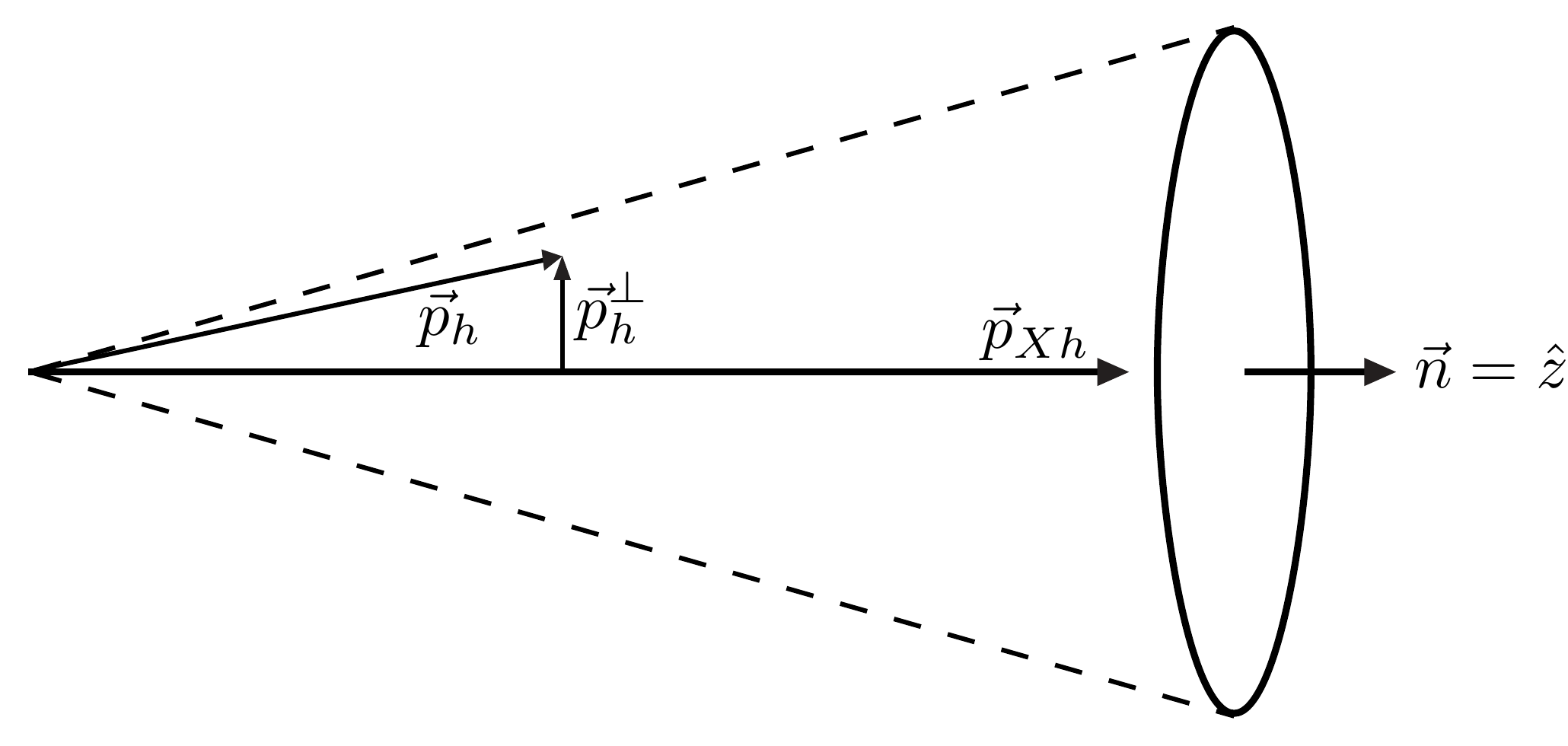}
\caption{Fragmentation of the hadron $h$ inside the jet $X h$.}
\label{fig:frag}
\end{figure}

The fragmenting jet functions (FJFs) $\cG_i^h(s,z)$ describe the fragmentation of a light parton $i$ to a light hadron $h$ within a jet originating from $i$, where in addition to the momentum fraction $z$, the invariant mass $s$ of the jet is measured~\cite{Procura:2009vm}. Here, we will consider generalized FJFs, which depend also on the transverse momentum of the hadron $p_{h\perp}^\mu \sim \sqrt{s}$ with respect to the jet axis, see fig.~\ref{fig:frag}. We could define the generalized gluon FJF with uncontracted indices, as we did for the gluon BF in \eq{B_def}. However, the new Lorentz structure only contributes at higher orders, especially for $e^+e^-$ collisions, due to the need of contracting indices with another gluon matrix element. 
We therefore restrict ourselves to the case with contracted Lorentz indices, for which only the magnitude $|\vec p_h^\perp|$ is relevant. The transverse momentum measurement can be described by inserting in the definition of the standard $\cG_i^h$ in refs.~\cite{Procura:2009vm,Liu:2010ng,Jain:2011xz} the additional function $\de^2(\vec p_\perp - \vec p_{h\perp}) = \de(\vps - \vphs)/\pi$.

The quark and gluon generalized FJFs are thus defined as 
\begin{align} \label{eq:G-def}
  \cG_{q,\bare}^h(s,z,\vps) &= 
  4(2\pi)^3 \!
\int\! \df^d p_h\, \de(p_h^2) \, \de(p_h^- - z k^-)\, \frac{1}{\pi} \de(\vps-\vphs)\,   \sum_X  \frac{1}{2 N_c}
  \nn \\ & \quad \times  
 \tr \Big[\frac{\bnslash}{2} \, 
   \Mae{0}{[\de(k^- -\bnP_n)\, \de^{d-2}(\vec \cP_n^\perp) \chi_n(0)]\, \de( k^+ - \hat{p}^+)} {X h}
  \Mae{X h}{\bar \chi_n(0)}{0} \Big]\,, \nn \\
 \cG_{g,\bare}^{h}(s,z,\vps) & = 
  - 4(2\pi)^3\, k^- \!\int\! \df^d p_h\, \de(p_h^2) \, \de(p_h^- - z  k^-)\, \frac{1}{\pi} \de(\vps-\vphs)\,  \sum_X   \frac{1}{(d-2) (N_c^2-1)}
  \nn \\ & \quad \times  
   \Mae{0}{ [\de(k^- -\bnP_n) \de^{d-2}(\vec \cP_n^\perp) \cB_{n\perp}^{\mu, a}(0)]\, \de(k^+ - \hat{p}^+)} {X h}
 \Mae{X h}{\cB_{n\perp\mu}^{a} (0)}{0} \,,
\end{align}
where $s = k^2 = k^- k^+$ is the invariant mass of the collinear radiation building up the jet (including the observed hadron).
The label momentum operators $\bnP$ and $\vec \cP_n^\perp$ act on the fields inside the square brackets, and the residual momentum operator $\hat p^+$ acts on the intermediate states. The state $ |X h \rangle = | X h (p_h)\rangle$ contains a hadron with momentum $p_h$, and a sum over the polarizations of $h$ is assumed.
At variance with ref.~\cite{Procura:2009vm}, we choose to combine the label and residual minus and perpendicular momentum components into continuous variables (see also ref.~\cite{Jain:2011xz}). 
In $\cG_q^h$ the trace is taken over color and Dirac indices, and the factor $1/(2 N_c)$, where $N_c = 3$ is the number of colors, comes from averaging over the color and spin of the parent parton. In $\cG_g^h$ an average over colors and the $(d-2)$ polarizations of the gluon is performed. 

In ref.~\cite{Procura:2009vm}, a simple replacement rule was obtained that allows one to obtain the factorization theorem, when the momentum fraction $z$ of a hadron in a jet is measured, from the corresponding inclusive case. Assuming cancellation of Glauber contributions, this can be extended to the situation where $|\vec p_h^\perp|$ is also measured, 
\begin{equation}\label{eq:jet-repl}
 J_i(s) \longrightarrow
\frac{1}{2\, (2 \pi)^3}\, \cG_i^h(s, z, \vec p_{h\, \perp}^{\;2} ,\mu)\, \df z\, \pi \df \vec p_{h\, \perp}^{\;2} \,.
\end{equation}
This equation is consistent with the fact that FJFs occur in \SCETa factorization theorems where the transverse momentum of collinear particles is much larger than that of ultra-soft momenta. Soft modes do not contribute for the same reason as in the beam function case. Therefore the observed hadron only recoils against the other collinear particles within that jet. The replacement rule in \eq{jet-repl} still holds in the case of two (or more) gluon FJFs, if the angles of $p_{h \perp}$ are not observed. In this case the second Lorentz structure does not contribute.

As a consequence of eq.~(\ref{eq:jet-repl}), the jet function $J_i$ and the FJF $\cG_i^h$ have the same renormalization, and thus the same anomalous dimension,
\begin{align}  \label{eq:G_ren}
  \cG_{i,\bare}^h (s,z, \vphs) & = \int_0^s\! \df s'\, Z_\cG^i(s - s',\mu)\, \cG_i^h(s',z, \vec p_{h\, \perp}^{\;2},\mu)
  \,, \qquad
  Z_\cG^i(s,\mu) =   Z_J^i(s,\mu)
  \,, \nn \\
  \mu \frac{\df}{\df \mu} \cG_i^h(s, z, \vphs, \mu) & = \int_0^s\! \df s'\, \ga_\cG^i(s-s',\mu)\,  \cG_i^h(s', z,  \vec p_{h\, \perp}^{\;2}, \mu)
  \,, \qquad
  \ga_\cG^i(s,\mu) =  \ga_J^i(s,\mu)
  \,,
\end{align}
where the parton type labelled by the index $i$ is not summed over. In analogy with the argument for the all-order renormalization structure in \subsec{ren_BF}, let us take as example the decay rate of the single-inclusive jet-like process $B \to (X \pi)_u \ell \nu$, where the leading-power factorization theorem has the structure~\cite{Procura:2009vm}
\begin{align} \label{eq:fact_ex}
 \df \Gamma = H\, \cG_u^\pi \otimes S~. 
\end{align}
The measurement of the hadron $\perp$-momentum affects only one function, namely $\cG_u^\pi$, the hard coefficient $H$ and the soft function $S$ do not change. Since the anomalous dimensions of all the functions of the RHS of \eq{fact_ex} must cancel, the $\perp$-momentum of the hadron cannot change the anomalous dimension of $\cG_i^h$. More intuitively, the integral over $\vphs$ is bound once $s$ is measured since $\vphs = p_h^- p_h^+ \leq k^- k^+ = s$. Therefore integrating over $\vphs$ cannot produce UV divergences.

\section{One-loop Matching}
\label{sec:oneloop}

In this section we discuss the matching of beam functions onto PDFs and fragmenting jet functions onto fragmentation functions. The one-loop matching coefficients are presented. We also provide the conversion relation from perpendicular-momentum to impact-parameter space. We compare to previous results for the matching coefficients in refs.~\cite{Mantry:2009qz,Mantry:2010mk} and discuss the disagreement.

\subsection{Generalized Beam Functions}\label{sec:gBF-matching}

In SCET the PDFs are defined as matrix elements of collinear fields with collinear proton states and are equivalent to the PDFs in QCD~\cite{Bauer:2002nz}. 
We illustrate this by giving the definition of the quark PDF in SCET and QCD respectively, 
\begin{align}
f_q^\text{SCET}(k^-/p^-, \mu)
&= \theta(k^-) \MAe{p_n(p^-)}{\bar\chi_n(0)\,
  \delta(k^- - \bnP_n)\,\frac{\bnslash}{2}\chi_n(0)}{p_n(p^-)}
\,, \\
f_q^\text{QCD}(k^-/p^-, \mu) &= \theta(k^-) \int\! \frac{\df y^+}{4\pi}\,
  e^{-\img k^- y^+/2}
  \MAe{p_n(p^-)}{\bar\psi \Bigl(y^+\frac{\bn}{2}\Bigr) \frac{\bnslash}{2}\,
   W_{\bn}\Bigl(y^+\frac{\bn}{2},0\Bigr) \psi(0)}{p_n(p^-)}
\,.\nn\end{align}
The Fourier transform over the light-cone separation $y^+$ of the two fields in QCD is replaced by a delta function of the conjugate large label momentum in SCET. The Wilson line in the QCD definition is absorbed into the SCET fields, see \eq{chiB}. In SCET the PDFs do not contain full QCD fields but collinear fields, which all belong to the same collinear sector. Each collinear sector is a boosted copy of QCD. Collinear fields may have zero-bin subtractions, but these are not present for the PDF, since it is insensitive to soft radiation. (QCD fields have no such subtraction.)

For $\{t, \vks\} \gg \lqcd^2$,  BFs can be related to the PDFs by performing an operator product expansion in $\lqcd^2/t$ and $\lqcd^2/\vks$, 
\begin{align} \label{eq:B_OPE2}
  B_q(s,x,\vks,\mu_B) 
  & = \sum_j \int_x^1 \frac{\df x'}{x'}
   \cI_{qj}\Big(t,\frac{x}{x'},\vks,\mu_B\Big) f_j(x',\mu_B)
   \bigg[1+ \ORd{\frac{\lqcd^2}{t}, \frac{\lqcd^2}{\vks}}\bigg]
  \,,
  \\
  B_g^{\mu\nu}(t,x,\vec k_\perp,\mu_B) 
  & = \sum_j \int_x^1 \frac{\df x'}{x'}
   \bigg[
   \frac{g_\perp^{\mu\nu}}{d-2}\, \cI_{gj}\Big(t,\frac{x}{x'},\vks,\mu_B\Big)  
   \nn \\ & \quad
    + \bigg(\frac{k_\perp^\mu k_\perp^\nu}{\vks} + \frac{g_2^{\mu\nu}}{2} \bigg)
   \tilde \cI_{gj}\Big(t,\frac{x}{x'},\vks,\mu_B\Big) \bigg]
   f_j(x',\mu_B)
   \bigg[1+ \ORd{\frac{\lqcd^2}{t}, \frac{\lqcd^2}{\vks}}\bigg]
  \,. \nn
\end{align}
where the index $j$ sums over parton flavors.
As stated in the introduction, we have assumed that $t$ and $\vks$ are parametrically of the same size and thus get ``integrated out" at the same time. The matching coefficients $\cI_{ij}$ and $\tilde \cI_{ij}$ will contain logarithms of $\vks/t$, which are small because of this assumption. 

We determine the matching coefficients in \eq{B_OPE2} by replacing the proton states in the BFs and PDFs by a collinear quark or a collinear gluon. At one loop, where at most one emission takes place, the invariant mass of the ISR is $r^2 = 0$. Eq.~\eqref{eq:kin} then becomes an equality, 
\begin{equation} \label{eq:B_rel}
\vks = \frac{(1-x)t}{x}
\,.\end{equation}
It is now straightforward to obtain the one-loop matching coefficients $\cI_{ij}$ from the calculation of the one-loop standard beam function in refs.~\cite{Stewart:2010qs,Berger:2010xi}, by including the extra perpendicular-momentum $\delta$-function in \eq{B_def} and using \eq{B_rel}. The one-loop calculation of the gluon BF with uncontracted Lorentz indices is given in \app{gfupdf}, from which we obtain $\tilde \cI_{ij}$.
The tree-level and one-loop matching coefficients are given by
\begin{align} \label{eq:Iresult}
\cI_{ij}^\zero (t,x,\vks,\mu_B) &= \frac{1}{\pi}\, \de_{ij}\, \de(t)\, \de(\vks) \de(1-x)
\,,\nn \\
  \tilde \cI_{gj}^\zero (t,x,\vks,\mu_B) &= 0
  \,, \nn \\
\cI_{ii}^\one(t,x,\vks,\mu_B)
& = \frac{\al_s(\mu_B) C_{ii}}{2\pi^2}\, \theta(x)\, \bigg\{
\frac{2}{\mu_B^2} \cL_1\Big(\frac{t}{\mu_B^2}\Big)\, \de(1-x)\, \de(\vks)
+ \frac{1}{\mu_B^2} \cL_0\Big(\frac{t}{\mu_B^2}\Big)\, P_{ii}(x)
\nn \\ & \quad 
\times \de\Big(\vks - \frac{(1-x) t}{x} \Big) + \de(t)\, \de(\vks) \Big[-P_{ii}(x) \ln x - \frac{\pi^2}{6} \de(1-x) + \cI_{ii}^{\de}(x) \Big] \bigg\}
\,, \nn \\
\cI_{ij}^\one(t,x,\vks,\mu_B)
& =  \frac{\al_s(\mu_B) C_{ij}}{2\pi^2}\, \theta(x)\, \bigg\{\Big[\frac{1}{\mu_B^2} \cL_0\Big(\frac{t}{\mu_B^2}\Big)\,\de\Big(\vks - \frac{(1-x)t}{x} \Big) 
 \nn \\ & \quad
+ \de(t)\, \de(\vks) \ln \frac{1-x}{x} \Big] P_{ij}(x) + \de(t)\, \de(\vks) \,\cI_{ij}^{\de}(x)  \bigg\} 
\,, \nn \\
  \tilde \cI_{gj}^\one (t,x,\vks,\mu_B) &=
-\frac{\al_s(\mu_B) C_{gj}}{\pi^2}\, \frac{1}{\mu_B^2} \cL_0\Big(\frac{t}{\mu_B^2}\Big)\, \theta(x)\theta(1-x) \frac{1-x}{x}
\de\Big( \vks - \frac{1-x}{x} t \Big)
\,.\end{align}
where the plus distributions $\cL_n$ are defined as
\begin{align} \label{eq:plusdef}
\cL_n(x)
&\equiv \biggl[ \frac{\theta(x) \ln^n x}{x}\biggr]_+
 = \lim_{\beta \to 0} \biggl[
  \frac{\theta(x- \beta)\ln^n x}{x} +
  \delta(x- \beta) \, \frac{\ln^{n+1}\!\beta}{n+1} \biggr]
\,,\end{align}
and satisfy the boundary condition $\int_0^1 \df x\, \cL_n(x) = 0$.
The expressions in \eq{Iresult} involve the color factors
\begin{equation} \label{eq:colorF}
  C_{qq} = C_{gq} = C_F\,, \quad
  C_{gg} = C_A\,, \quad
  C_{qg} = T_F\,,
\end{equation}
the splitting functions
\begin{align} \label{eq:split}
  P_{qq}(x) &= (1+x^2)\, \cL_0(1-x)\, ,\; &
  P_{qg}(x) &= \theta(1-x)\, [x^2+(1-x)^2]
  \,, \nn \\
  P_{gg}(x)
  &= 2x\, \cL_0(1-x) + 2\,\theta(1-x)\Bigl[\frac{1-x}{x} +  x(1-x)\Bigr]\, ,\; &
  P_{gq}(x) &= \theta(1-x) \,\frac{1+(1-x)^2}{x}
  \,,
\end{align}
[where $P_{qq}$ is unconventionally defined without the $3/2\,\de(1\!-\!x)$ term to simplify \eq{Iresult}] and
\begin{align} \label{eq:cI}
\cI_{qq}^{\de}(x) &= (1+x^2) \cL_1(1-x) + \theta(1-x)\,(1-x)\,,&
\cI_{qg}^{\de}(x) &= 2\theta(1-x)\,x(1-x) \,, \nn \\
\cI_{gg}^{\de}(x) &= \frac{2(1-x+x^2)^2}{x}\cL_1(1-x)\,, &
\cI_{gq}^{\de}(x) &= \theta(1-x)\,x\,.
\end{align}
Note that the cancellation of IR divergences in the one-loop matching of the standard BFs~\cite{Stewart:2010qs}, immediately carries over to the generalized BFs as well.

\subsection{Comparison with Matching Coefficients of iBFs} \label{sec:sonny}

The conversion of our results to impact-parameter space is needed to compare with the matching coefficients for iBFs in refs.~\cite{Mantry:2009qz,Mantry:2010mk}. Here we give the necessary equations both for CDR and CDR${}_2$ schemes. We will show that for our calculation CDR and CDR${}_2$ yield the same result in impact-parameter space (even though CDR is not suitable for momentum space, see \subsec{HVvsCDR}). 

For the two Lorentz structures in \eq{Lor2} the transformation to the impact parameter $\vec y_\perp$ in CDR is given by
\begin{align} \label{eq:schemesip}
 & \int\! \df^{d-2} \vec k_\perp\, e^{\img  \vec y_\perp \cdot\, \vec k_\perp}\, \frac{g_\perp^{\mu\nu}}{d-2}\, \de^{d-2}(\vec k_\perp - \vec \cP_{n\perp}) 
 = \frac{g_\perp^{\mu\nu}}{d-2}\, {}_0F_1\Big(1-\eps, -\frac{\vec y_\perp^{\,2} \vcps}{4}\Big)
 \,, \nn \\
 &\int\! \df^{d-2} \vec k_\perp\, e^{\img  \vec y_\perp \cdot\, \vec k_\perp}\, \Big(\frac{k_\perp^\mu k_\perp^\nu}{\vks} + \frac{g_\perp^{\mu\nu}}{d-2}\Big)\, \de^{d-2}(\vec k_\perp - \vec \cP_{n\perp}) 
 \nn \\ & \qquad
 = \Big(\frac{y_\perp^\mu y_\perp^\nu}{\vec y_\perp^{\,2}} + \frac{g_\perp^{\mu\nu}}{d-2}\Big) \frac{1}{(1-\eps)(2-\eps)} \times \Big ( -\frac{\vec y_\perp^{\,2} \vcps}{4} \Big ) {}_0F_1\Big(3-\eps, -\frac{\vec y_\perp^{\,2} \vcps}{4}\Big)
\,,\end{align}
with $d=4-2\eps$. The corresponding expressions for CDR${}_2$ follow directly by taking all perpendicular quantities in two dimensions and setting $\eps=0$ in these equations [except for the $g_\perp^{\mu\nu}/(d-2)$ on the first line, which is just an overall factor]. The conversion needed for the quark beam function is the same as that of the $g_\perp^{\mu\nu}$ structure. To show how these results were obtained, we present the derivation of the first line of \eq{schemesip}:
\begin{align}
\int\! \df^{d-2} \vec k_\perp\, e^{\img  \vec y_\perp \cdot\, \vec k_\perp}\, \de^{d-2}(\vec k_\perp - \vec \cP_{n\perp}) 
&= \int\! \df^{d-2}  \vec k_{\perp}\,e^{\img \vec y_\perp \cdot\, \vec k_\perp} \, \frac{2\Ga(\frac{d}{2})}{(d-2)\pi^{d/2-1}} |{\vec k}_{\perp}|^{4-d}\, \de(\vks - \vcps) \nn \\ &
= \frac{\Ga(\frac{d}{2})}{(d-2)\pi^{d/2-1}} \int\! \df^{d-3} \Omega\, \df \vks \,e^{i |\vec y_{\perp}| |\vec k_\perp| \cos\phi} \, \de(\vks - \vcps) \nn \\ &
= {}_0F_1\Big(1-\eps, -\frac{\vec y_\perp^{\,2} \vcps}{4}\Big)\,,
\end{align}
The angular integral was performed using
\begin{align}
\int\! \df^{d-3} \Omega = \frac{2 \pi^{(d-3)/2}}{\Gamma\big(\frac{d-3}{2}\big)} \int_0^\pi\! \df \phi\, \sin^{d-4} \phi \,,
\end{align}
where $\phi$ is the angle between $\vec k_\perp$ and $\vec y_\perp$.

In \subsec{GBF} we showed that the renormalization of the BF does not affect $\vec k_\perp$, so the UV divergences are multiplied by $\de(\vec k_\perp)$. Since the PDF does not depend on $\vec k_\perp$ (by definition), the IR divergences in the BF are multiplied by $\de(\vec k_\perp)$ as well. Therefore any $\ord{\eps}$ pieces that may appear in \eq{schemesip} due to the difference between CDR and CDR${}_2$ never get enhanced by $1/\eps$ divergences,
\begin{align}
 {}_0F_1(1-\eps,0) &= 1 = {}_0F_1(1,0)\,, \qquad
 {}_0F_1\Big(1-\eps, -\frac{\vec y_\perp^{\,2} \vcps}{4}\Big) = {}_0F_1\Big(1, -\frac{\vec y_\perp^{\,2} \vec \cP_{n\perp}^{\,2}}{4}\Big) + \ord{\eps}\,.
\end{align}

Using \eq{schemesip}, it is now straightforward to compare our results with the matching coefficients for the iBFs in refs.~\cite{Mantry:2009qz,Mantry:2010mk}. We agree with $\cI_{gg}$, $\cI_{qq}$ and the matching coefficient for the new $k_\perp$-dependent Lorentz structure in \eq{Iresult}. We disagree with $\cI_{gq}$ and $\cI_{qg}$ and discuss what we believe is the problem in refs.~\cite{Mantry:2009qz,Mantry:2010mk}. Their calculation relies on the fact that in pure CDR the partonic $f_{i/j}$ is
\begin{align} \label{eq:fqgone}
 f_{i/j}(x,\mu) = \de_{ij} \de(1-x) - \frac{1}{\eps}\,\frac{\alpha_s(\mu) C_{ij}}{2\pi}\, \theta(x) P_{ij}(x)
\,,\end{align}
implying that $\cI_{ij}$ is simply the finite part of the one-loop beam function,
\begin{align} \label{eq:onematch}
\Big[ B_{i/j}^\one\Big]_\text{finite} &= \Big[\cI_{ii}^\zero \otimes f_{i/j}^\one + \cI_{ij}^\one \otimes f_{j/j}^\zero\Big]_\text{finite}
  = \cI_{ij}^{\one}
\,.\end{align}
However, the definition of the PDF, which yields \eq{fqgone}, requires one to average over the spins of the incoming parton. They do not do this correctly for $B_{g/q}$ and $B_{q/g}$, where they sum over the spins and then divide by $d-2$ (rather than 2) for $B_{g/q}$ and $2$ (rather than $d-2$) for $B_{q/g}$. This incorrect averaging leads to a modification of the matching in \eq{onematch}. For example, for $B_{q/g}$
\begin{align}
 \Big[B_{q/g}^\one\Big]_\text{finite} &= \Big[\cI_{qq}^\zero \otimes f_{q/g}^\one \times (1-\eps) + \cI_{qg}^\one \otimes f_{g/g}^\zero\Big]_\text{finite}
  = \cI_{qg}^\one + \frac{\alpha_s(\mu) T_F}{2\pi}\, \theta(x) P_{qg}(x)
\,.\end{align}
Including this additional term, their result agrees with our calculation. Note that since in ref.~\cite{Stewart:2010qs} and in \app{qfupdf} of the present paper the PDFs are explicitly calculated, these derivations are not prone to this problem. A similar issue occurs for $\cI_{gq}$, but there in addition the sign of the $\de(t) \ln[(1-x)/x]  P_{gq}(x)$ term in $\cI_{gq}$ does not agree with the calculations in ref.~\cite{Berger:2010xi} and our \app{gfupdf}. We were unable to identify the origin of what we believe is a sign error in Ref.~\cite{Mantry:2010mk}.

\subsection{Fully-Unintegrated Fragmentation Functions}

The FFs in SCET, which are defined in terms of collinear fields and collinear intermediate states $|X h \rangle$~\cite{Procura:2009vm,Jain:2011xz}, are equivalent to the standard QCD FFs that appear in factorization theorems at leading power, since the sum over the intermediate states is dominated by jet-like configurations~\cite{Collins:1989gx}. As for PDFs, the zero-bin contributions cancel against each other in the sum over the graphs at each order, because these functions are insensitive to the scale associated with the ultra-soft radiation accompanying the parton~\cite{Jain:2011xz}. (This is not the case for FJFs, where the zero-bin contributions do not completely cancel each other.)

The matching of the FJFs onto standard FFs corresponds to an operator product expansion analogous to the case of the BFs. Our calculation at one loop in ref.~\cite{Jain:2011xz} represents an explicit check that the additional measurement on the jet invariant mass does not spoil the cancellation of the IR divergences between FJFs and FFs.

Starting from
\begin{align} \label{eq:G_OPE}
  \cG_i^h(s,z,\vphs,\mu_J) 
  & = \sum_j \int_z^1 \frac{\df z'}{z'}
   \cJ_{ij}\Big(s,\frac{z}{z'},\vphs,\mu_J\Big) D_j^h(z',\mu_J)
   \bigg[1+ \ORd{\frac{\lqcd^2}{s},\frac{\lqcd^2}{\vphs}}\bigg]
  \,,
\end{align}
the matching is performed by replacing the observed hadron in the FJF and in the FFs by a collinear quark or a collinear gluon. With at most one real emission, the perpendicular momentum $\vec p_\perp^{\,2}$ of this parton is completely fixed in terms of $s$ and $z$
\begin{equation} \label{eq:G_rel}
 \vps = z(1-z) s
\end{equation}
which follows from momentum conservation and the on-shell condition $p^2 = 0$. 
As for the BFs, we can now straightforwardly obtain the matching coefficients from the calculation of the one-loop standard fragmenting jet function in ref.~\cite{Jain:2011xz}, by adding the delta function for the perpendicular momentum. There the quark fragmenting jet function was calculated using a gluon mass $m$ as IR regulator. The gluon mass would in principle modify \eq{G_rel}. However, at the end of the calculation the limit $m \to 0$ is taken to isolate the IR divergences. It is safe to take this limit ahead of time in \eq{G_rel} since it does not modify the result for these divergences.

We find that the one-loop matching coefficients are given by
\begin{align} \label{eq:Jresult}
\frac{\cJ_{ii}^\one(s,z,\vphs,\mu_J)}{2(2\pi)^3}
& = \frac{\al_s(\mu_J) C_{ii}}{2\pi^2}\, \theta(z)\, \bigg\{
\frac{2}{\mu_J^2} \cL_1\Big(\frac{s}{\mu_J^2}\Big)\, \de(1-z)\, \de(\vphs)
+ \frac{1}{\mu_J^2} \cL_0\Big(\frac{s}{\mu_J^2}\Big) P_{ii}(z)
\nn \\ & \quad 
\times \de[\vphs - z(1-z) s] + \de(s)\, \de(\vphs) \Big[P_{ii}(z) \ln z - \frac{\pi^2}{6} \de(1-z) + \cI_{ii}^{\de}(z) \Big] \bigg\}
\,, \nn \\
\frac{\cJ_{ij}^\one(s,z,\vphs,\mu_J)}{2(2\pi)^3}
& =  \frac{\al_s(\mu_J) C_{ji}}{2\pi^2}\, \theta(z)\, \bigg\{\Big[\frac{1}{\mu_J^2} \cL_0\Big(\frac{s}{\mu_J^2}\Big)\,\de(\vphs - z(1-z) s) 
 \nn \\ & \quad
+ \de(s)\, \de(\vphs) \ln{(z(1-z))} \Big] P_{ji}(z) + \de(s)\, \de(\vphs) \,\cI_{ji}^{\de}(z)  \bigg\} 
\,,
\end{align}
where $C_{ij}$, $P_{ij}(z)$ and $\cI_{ij}^\de(z)$ were given in eqs.~(\ref{eq:colorF}), (\ref{eq:split}) and (\ref{eq:cI}).

\section{Examples of Factorization Theorems}
\label{sec:factth}

This section is devoted to illustrative examples of leading-power factorization theorems involving generalized BFs and FJFs. The first one concerns the transverse momentum $Q_\perp$-distribution of a Drell-Yan lepton pair at perturbative values of $Q_\perp$ with a veto on hard central jets. This factorization formula can be straightforwardly extended to the case of Higgs production through gluon fusion ($gg \to H$) and Higgs Strahlung $q \bar{q} \to H V$. The factorization theorem for $gg \to H$ was used in \subsec{ren_BF} to discuss the renormalization of BFs.
The purpose of this paper is to describe the general picture associated with BFs, which suggests that they will play a role in describing transverse momentum distributions for events with any number of energetic well-separated jets, \eg $ p p \to W/Z + n~{\rm jets}$, once open factorization issues concerning for example the contribution from Glauber modes are settled. 

As a paradigm of factorization theorems involving FJFs, we will discuss the $p_\perp$-distribution of the observed hadron in $e^+ e^- \to X h$, where a cut on the thrust event shape is applied to restrict to dijet final-state configurations. Such a cut on thrust is utilized by the Belle collaboration to study light-quark fragmentation in their on-resonance data, in order to remove the large $b$-quark background. Here we consider the case of spin-averaged fragmentation. When polarization and azimuthal correlations between hadrons in back-to-back jets are taken into account, our framework can be applied to study Belle data on polarized light di-hadron fragmentation in opposite hemispheres $e^+ e^- \to {\rm dijets} + 2 h$.

\subsection{Transverse Momentum Distribution in $pp \to \ell^+ \ell^- + 0$ jets}
\label{subsec:factth_DY}

We start our discussion by considering Drell-Yan (DY) production, $pp\to \ell^+\ell^-$, of lepton pairs with a large invariant mass $Q$. Factorization at leading power in $\lqcd/Q$ has been discussed by Collins, Soper and Sterman, for any value of the transverse momentum of the lepton pair $Q_\perp$ with respect to the beam axis, namely both cases $Q_\perp \sim Q$ and $Q_\perp  \ll Q$~\cite{Collins:1984kg}. 
Since $\vec{Q}_\perp$ equals the sum of the transverse momenta of the colliding partons entering the hard subprocess, the DY cross section is directly sensitive to the partonic transverse momenta. 

The rigorous proof of the cancellation of contributions from Glaubers modes, which amount to initial-state hadron-hadron interactions taking place before the annihilation, that would spoil factorization~\cite{Bodwin:1981fv}, has been achieved in ref.~\cite{Collins:1988ig} for the case of inclusive DY. Ref.~\cite{Stewart:2009yx} discusses an extension to the isolated DY process, $p p \to \ell^+ \ell^- + 0$ jets, where a central jet veto is imposed, restricting energetic ISR to be close to the beam axis. In this framework, the colliding partons are far from threshold (as in the inclusive case) and the collinear radiation is described by beam functions. Here we consider a more differential case, combining this with the study of transverse momentum distributions. We impose the central jet veto through a cut on the beam thrust event shape $\Tcm$, defined in the hadronic center-of-mass frame as~\cite{Berger:2010xi}
\begin{equation} \label{eq:TauBcm}
\Tcm
= \sum_k\, \abs{\vec p_k^\perp}\, e^{-\abs{\eta_k}}
\,.\end{equation}
The sum on $k$ runs over all particles in the final state, except for the signal leptons, where $\vec p_k^\perp$ and $\eta_k$ denote the transverse momentum and rapidity of the particle with respect to the beam axis. The jet veto, $\Tcm \leq \Tcm^\cut \ll Q$, leads to large logarithms $\al_s^n \ln^m \Tcm^\cut/Q$ (with $m\leq2n$) in the cross section. 

If $\lqcd \ll Q_\perp \simeq \sqrt{\Tcm\, Q} \ll Q$,
we can write the following leading-power factorization formula following ref.~\cite{Stewart:2009yx} 
\begin{align} \label{eq:DY_fact}
\frac{\df \sigma}{\df \Tcm\, \df \vec Q_{\perp}^{\;2}}
&= \sigma_0 \sum_{i j}\, H_{ij}(Q^2, \mu)  \int\!\df Y \!\int\!\df t_a\, \df t_b\, 
S_B^{ij}\Bigl(\Tcm \!-\! \frac{e^{-Y} t_a \!+\! e^Y t_b}{Q}, \mu\Bigr)
\int\! \frac{\df \phi}{2\pi} \int\! \df \vec k_{a\perp}^{\;2}\, \df \vec k_{b\perp}^{\;2} \pi^2
\nn\\ &\quad \times
\de\big[\vec Q_{\perp}^{\;2} \!-\! \big(\vec k_{a\perp}^{\;2} \!+\! 2|\vec k_a^\perp||\vec k_b^\perp| \cos \phi \!+\! \vec k_{b\perp}^{\;2}\big)\big]\,
B_i(t_a, z_a, \vec k_{a\perp}^{\;2}, \mu)\, B_j(t_b, z_b, \vec k_{b\perp}^{\;2},\mu)
\,,\end{align}
where $\si_0$ is the Born cross section and $\phi$ is the angle between $\vec k_a^\perp$ and $\vec k_b^\perp$. The momentum fractions are given by
\begin{equation}
x_a = \frac{Q}{\ECM}\,e^{Y}
\,,\qquad
x_b = \frac{Q}{\ECM}\,e^{-Y}
\,,\end{equation}
with $\ECM$ being the center-of-mass energy and $Y$ the total rapidity of the leptons. The sum in \eq{DY_fact} extends over the various quark flavors without mixed terms, \ie $i j = \{u \bar{u}, 
\bar{u} u, d \bar{d} , \dots \}$. 

The hard function $H_{q\bar q}$ encodes the virtual effects at the hard scale $Q$ due to the underlying hard process. It only depends on the large momentum components and not on the perpendicular momenta. The ultra-soft radiation is described by the soft function $S_B^{q\bar q}$, and the contribution of the soft radiation to $Q_\perp$ is power suppressed, due to our hierarchy of scales. Therefore $H_{q\bar q}$ and the soft function $S_B^{q\bar q}$ are the same as in ref.~\cite{Stewart:2009yx}.
Only the BFs account for the recoil of the energetic initial-state radiation against the final-state leptons. 

Eq.~\eqref{eq:DY_fact} and the one-loop matching coefficients in \sec{oneloop} provide all the ingredients needed to analyze up to NNLL the DY cross section in terms of $\Tcm$ and $Q_\perp$, which can be tested for example against LHC data where the effect of resummation of large logs is expected to be important due to the large separation of scales.
Because of our assumption $\vec Q_{\perp}^{\;2} \sim \Tcm\, Q$, the logarithms of $\Tcm\, Q/\vec Q_{\perp}^{\;2}$ in the cross section are not large. These are generated by the corresponding logarithms of $t/\vks$ in the beam functions, through the convolutions in \eq{DY_fact}.

Our approach provides two handles on the ISR, by simultaneously measuring its transverse momentum and beam thrust. This can be useful in the discovery and interpretation of new physics according to the authors of ref.~\cite{Krohn:2011zp}. For the example of SUSY di-squark (and di-gluino) production at threshold, their method leads to the determination of the sparticle mass. This entails a boost along the (beam) $z$-axis to the frame where the net $p_z$ of the final state radiation is zero. They determine the boost parameter from measurements on the final-state radiation, but this may also be achieved using beam thrust \cite{Stewart:2010tn}, which has the advantage of being unaffected by the presence of invisible final-state particles. In addition, their approach relies on the determination of the transverse momentum of the ISR. Our paper provides the tools for a reliable calculation of the cross section differential in beam thrust and the transverse momentum of the ISR, which is the input of their method.

\subsection{Measuring the Full Hadron Momentum in $e^+e^- \to {\rm dijet} + h$}

In ref.~\cite{Jain:2011xz}, the spin-averaged fragmentation of a light hadron was studied in the process $e^+ e^- \to {\rm dijet} + h$. 
Experimentally \cite{Seidl:2008xc}, the dijet limit is imposed by a cut on the thrust event shape~\cite{Farhi:1977sg}
\begin{equation} \label{eq:thrust}
  T = \max{}_{\hat t}\; \frac{\sum_i |\hat t \sdt {\vec p}_i|}{\sum_i |{\vec p}_i|} 
  \,,
\end{equation}
which we will use as well. In \eq{thrust}, the sum runs over all particles in the final state and the value of $\hat t$ that maximizes $T$ is called the thrust axis, which in the two-jet limit is along the direction of the jets. We will use the variable $\tau=1-T$, where $\tau$ close to 0 corresponds to configurations with two narrow, pencil-like, back-to-back jets. The maximum value of $\tau=1/2$ instead corresponds to a spherically symmetric event. The cut on thrust, $\tau \leq \tau^\cut \ll 1$, leads to double logarithms $\al_s^n \ln^m \tau$ ($m\leq2n$) in the cross section, that were resummed up to NNLL in ref.~\cite{Jain:2011xz}. Here, we consider the situation where, in addition, the transverse momentum of the hadron $p_h^\perp$ with respect to the thrust axis is measured. For intermediate values of $p_{h \perp}$, namely for $\lqcd \ll p_{h \perp} \sim \sqrt{\tau} \ECM \ll \ECM$ (in accordance with the \SCETa power counting) the following factorization formula holds
\begin{align} \label{eq:frag_fact}
  \frac{\df \si}{\df \tau\, \df z\, \df \vphs} &= 
\sum_q \frac{\sigma_0^q}{(4 \pi)^2}\, H(Q^2, \mu) \int\! \df s_a\, \df s_b\, Q\, S_\tau\Big(Q\tau - \frac{s_a+s_b}{Q^2},\mu \Big) 
\nn\\ &\quad \times
 \Big[\cG_q^h(s_a, z, \vphs, \mu)\, J_{\bar{q}}(s_b,\mu) + J_q(s_a, \mu)\, \cG_{\bar{q}}^h(s_b,z, \vphs, \mu)\Big]
  \,,
\end{align}
where $\si_0$ is the Born cross section.
Since the hadron is produced by the energetic radiation inside a jet, assuming cancellation of Glauber contributions, the only modification required to the right-hand side of the factorization theorem in ref.~\cite{Jain:2011xz} is the replacement of the standard FJFs by the generalized FJFs.
The hard function $H$ encodes virtual effects from the hard process at the hard scale $\ECM$, the soft function $S_\tau$ describes the contribution to thrust due to ultra-soft emissions, and the jet function $J$ is associated with the energetic radiation in the jet which does not contain the detected hadron. All these functions are the same as in ref.~\cite{Jain:2011xz}. Together with the matching coefficients in \sec{oneloop} we provide all the ingredients to determine the cross section in \eq{frag_fact} at NNLL order. Since $p_{h \perp} \sim \sqrt{\tau}Q$, the logarithms of $\sqrt{\tau}Q/p_{h \perp}$ in the cross section are not large.

\section{Conclusions}
\label{sec:concl}

In this paper we defined and studied the generalized BFs and FJFs.
The former describe the distribution in the full four momentum $k^\mu$ of a colliding parton taken out of the incoming hadron, for perturbative values of $k_\perp^\mu$. We considered $t= -k^+k^- \sim \vks$ which avoids introducing an additional hierarchy between these two scales. Generalized BFs are relevant for factorization theorems for exclusive $N$-jet processes, when the measurement is sensitive to the transverse momentum of the ISR. We show explicitly that a proper definition of generalized BFs requires ultra-soft zero-bin subtractions. We have argued that the renormalization of the generalized BF is the same as that of the jet function, and discussed an important subtlety in defining and calculating perpendicular momentum dependence using dimensional regularization. We have calculated the one-loop matching of BFs onto standard PDFs and corrected the non-diagonal coefficients $\cI_{gq}$ and $\cI_{qg}$ previously worked out in ref.~\cite{Mantry:2010mk}. 

We similarly discussed the generalized FJFs, which describe the full momentum $p_h^\mu$ dependence of a hadron $h$ fragmenting from a parton with virtuality $s$. The perturbative regime $s \sim \vphs \gg \lqcd^2$ is assumed. An example for which we give the factorization theorem involving generalized FJFs is $e^+e^- \to \text{dijet}+h$, where a cut on the thrust event shape is used to impose the dijet limit and the full hadron momentum is measured. The Belle collaboration employs such a cut on thrust to remove $B$-meson events from the data sample on the $\Upsilon(4S)$ resonance and study light-quark fragmentation \cite{Seidl:2008xc}. The extension of our framework to include spin correlations, would allow one, for example, to study the Collins effect \cite{Collins:1992kk}. We showed that the renormalization of the FJF is equal to that of the jet function and obtained the one-loop Wilson coefficients for matching FJFs onto standard fragmentation functions.

\begin{acknowledgments}
We thank Iain Stewart, Aneesh Manohar and Duff Neill for useful discussions. We thank Iain Stewart and Frank Tackmann for comments on this manuscript.
A.J.~is supported by DOE grant, 22645.1.1110173. 
M.P.~acknowledges support  by the ``Innovations- und Kooperationsprojekt C-13'' of the Schweizerische Universit\"atskonferenz SUK/CRUS and by the Swiss National Science Foundation. 
W.W.~is supported by DOE grant DE-FG02-90ER40546. W.W.~acknowledges  support by the National Science Foundation under Grant No.~1066293 and thanks the Aspen Center for Physics for hospitality.
We thank the Department of Energy's Institute for Nuclear Theory at the University of Washington for its hospitality and partial support during the completion of this work.
\end{acknowledgments}

\appendix

\section{One-Loop Quark BF with IR and Rapidity Regulators}
\label{app:qfupdf}


We present detailed results for the quark generalized BF calculation at one loop, showing the structure of the IR and UV divergences, the zero-bin contribution and the cancellation of the rapidity divergences. To ensure an unambiguous calculation, we regulate the IR, UV and small rapidity region in the integral with distinct regulators. We use dimensional regularization for UV divergences, for IR divergences we use a gluon mass (or a quark mass for the mixing contribution) and to regulate rapidity divergences we use the $\delta$-regulator \cite{Chiu:2009yz}. Rapidity divergences arise in a loop-integral with eikonal propagators when $k^+k^-$ is fixed but either $k^+$ or $k^-$ goes to zero, see for example~\cite{Chiu:2011qc}. The $\de$-regulator of ref.~\cite{Chiu:2009yz} takes care of such divergences, however, historically it has been thought of as an IR regulator. In ref.~\cite{Jain:2011xz} we calculated the fragmenting jet functions and fragmentation functions employing this same choice of regulators. We find that the UV divergences extracted from this calculation of the generalized BF are the same as those of the jet function and the standard beam function~\cite{Stewart:2010qs,Stewart:2009yx}. We will show that only after an ultra-soft zero-bin subtraction the rapidity divergences are cancelled in the BF and the IR divergences match with those of the PDF, as expected. Consequently we find that the (quark) BF, without zero-bin subtractions, contains rapidity divergences.

\begin{figure}[t]
\centering
\includegraphics[width=\textwidth]{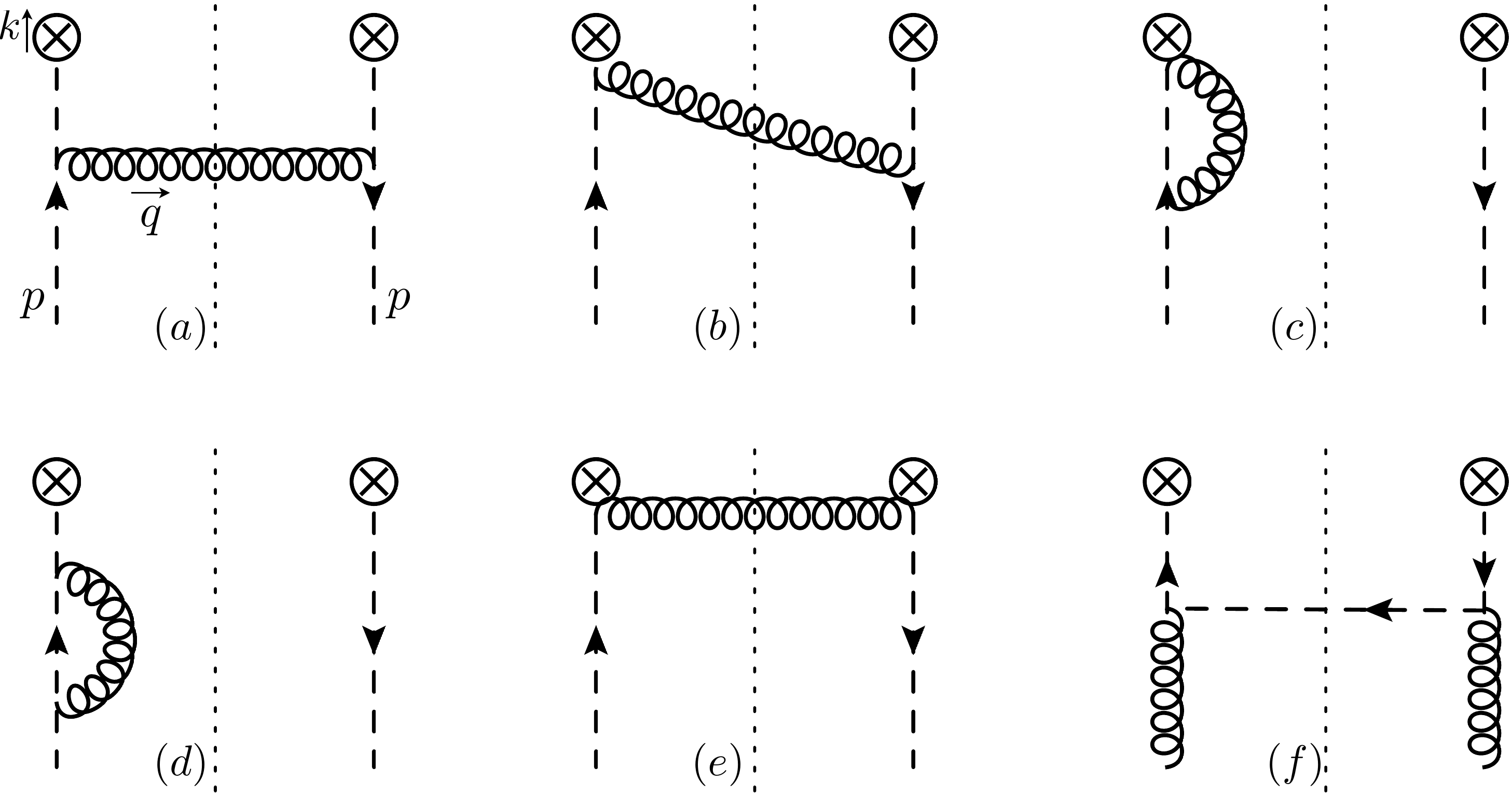}
\caption{Feynman graphs contributing to the quark BF at one loop. Graphs (b) and (c) have symmetric counterparts that are not shown but are included in their computation. Graph (e) vanishes in the Feynman gauge.
}
\label{fig:qfupdf}
\end{figure}

The graphs contributing to the quark BF at one loop are shown in \fig{qfupdf}. Including the symmetric counterparts for (b) and (c), the result for each graph is:
\begin{align}
B_a &=  \frac{\alpha_s(\mu) C_F}{2\pi^2}\,\de\Big( \vks - \frac{1-x}{x} t \Big) \theta(x) \theta(1-x) (1-x)  \bigg\{
\frac{1}{\mu^2}\cL_0\Big(\frac{t}{\mu^2}\Big) +\,  \delta (t) \Big[\ln \frac{\mu^2(1-x)}{m^2 x^2} -1\Big] \bigg\}
\, , \nn \\
B_b  &=   \frac{2\alpha_s(\mu) C_F }{2\pi^2}\, \de\Big( \vks - \frac{1-x}{x} t \Big)\, \theta(x) \bigg \{ 
\frac{1}{\mu^2}\cL_0\big(\frac{t}{\mu ^2}\big)  \Big[x \cL_0(1-x) - \delta (1-x) \ln \frac{\de}{k^-} \Big ]+
  \nn \\ & \quad
\delta (t)\Big[x \cL_1(1-x)+ x \cL_0(1-x)  \ln \frac{\mu ^2}{m^2 x^2} 
- \delta (1-x) \Big( \ln \frac{\mu^2}{m^2} \ln \frac{\de}{k^-} + \frac{1}{2}\ln ^2\frac{\de}{k^-}+\frac{\pi ^2 }{6} \Big) \Big]  \bigg \} \, ,\nn \\
B_c &=  \frac{2\alpha_s(\mu) C_F}{2\pi^2} \delta (t) \delta (1-x) \de(\vks) \bigg[
\Big(\frac{1}{\eps} + \ln \frac{\mu^2}{m^2} \Big)\Big(1+ \ln \frac{\de}{k^-}\Big) +1-\frac{\pi ^2}{6} \bigg]
\,. 
\end{align}
Graphs (b) and (c) have also non-vanishing zero-bin contributions:
\begin{align}
B_{b,0} 
&= \frac{2\alpha_s(\mu) C_F }{2\pi^2}\, \delta (1-x) \de(\vks) \bigg[
-\frac{1}{\mu^2}\cL_1\Big(\frac{t}{\mu^2}\Big)
+ \frac{1}{\mu^2}\cL_0\Big(\frac{t}{\mu^2}\Big) \Big(  \frac{1}{\epsilon }-\ln \frac{\delta }{k^-} \Big) 
\\ & \quad
 + \de(t) \Big( \frac{1}{\epsilon } \ln \frac{\delta }{k^-}-\frac{2}{\epsilon } \ln \frac{\delta }{\mu }-\frac{1}{2} \ln ^2\frac{\delta }{k^-}+2 \ln ^2\frac{\delta }{\mu }+\frac{\pi ^2}{6}  \Big ) \bigg] \,, \nn \\
B_{c,0} &= \frac{2\alpha_s(\mu) C_F }{2\pi^2}\, \delta (t) \delta (1-x) \de(\vks) \bigg( 
 -\frac{1}{\epsilon ^2}-2 \ln ^2\frac{\delta }{\mu}+\frac{2}{\epsilon } \ln \frac{\delta }{\mu}-\frac{\pi ^2}{4}  \bigg) \, .\nn
\end{align}
The wave function renormalization gives
\begin{align}
B_{d} = \frac{\alpha_s(\mu) C_F}{2\pi^2} \delta (t) \delta(1-x) \de(\vec k_\perp^2) \bigg(
-\frac{1}{2 \epsilon }-\frac{1}{2}\ln \frac{\mu ^2}{m^2}+\frac{1}{4} \bigg) \, .
\end{align}
Graph (e) vanishes in Feynman gauge.
The sum of these diagrams yields $B_{q/q}$,
\begin{align}
B_{q/q}(s,x,\vec k_\perp^2, \mu ) &= B_a + B_b + B_c +B_d - B_{b,0} - B_{c,0} \\
& =  \frac{\alpha_s(\mu) C_F }{2\pi^2} \de\Big( \vec k_\perp^2 - \frac{1-x}{x} t \Big)\, \theta(x) \bigg\{ 
\de(1-x) \bigg[ \frac{2}{\epsilon ^2}\, \delta (t)-\frac{2}{\epsilon }\,\frac{1}{\mu^2}\cL_0\Big(\frac{t}{\mu^2}\Big)+\frac{3}{2 \epsilon }\, \delta (t) \bigg] \nn \\
&\quad + \de(t)\ln\frac{\mu^2}{m^2}\, \Big[2 x \cL_0(1-x)+\frac{3}{2} \delta (1-x) + \theta(1-x)(1-x) \Big]   \nn \\
& \quad + \frac{2}{\mu^2}\cL_1\Big(\frac{t}{\mu^2}\Big)\, \de(1-x) + \frac{1}{\mu^2}\cL_0\Big(\frac{t}{\mu^2}\Big) \big[2x \cL_0(1-x) + \theta(1-x) (1-x) \big ] \nn \\
&\quad + \de(t) \bigg[ 2 x \cL_1(1\!-\!x) +\Big(\frac{9}{4}-\frac{\pi^2}{2}\Big) \delta (1\!-\!x)
\nn \\
 & \quad+\theta(1-x) \Big((1\!-\!x) \ln \frac{1\!-\!x}{x^2} - \frac{4x \ln x}{1\!-\!x}  -(1\!-\!x)\Big) \bigg] \bigg\} \, ,
\nn 
\end{align}
where the first line of the second equality contains the UV divergences, the second line the IR divergences that match with those of the PDFs (which are given below) and the last two lines show the finite contribution from BFs to the matching. As expected, the $\ln\de$ terms cancelled out after including zero-bin subtractions.

Graph (f) contributes to the mixing term in the matching\footnote{Here $m$ is quark mass rather than the gluon mass, but this neither affects the renormalization nor the matching.},
\begin{align}
B_f =  \frac{\alpha_s(\mu) T_F}{2\pi^2} \,\de\Big( \vks - \frac{1-x}{x} t \Big)\theta(x) P_{qg}(x) \bigg[\frac{1}{\mu^2}\cL_0\Big(\frac{t}{\mu^2}\Big) + \de(t) \ln \frac{\mu^2  (1-x)}{m^2 x}- \de(t) \bigg].
\end{align}
with the splitting function $P_{qg}$ given in \eq{split}. The crossed version of graph (f) does not contribute for positive values of $x$.

If the ultra-soft zero-bin subtractions are not performed, the $\ln \de$ will remain in the final answer, indicating the presence of rapidity divergences. If they are interpreted as IR divergences (as is typically done) then the BFs cannot be matched onto PDFs, since the IR divergences of the BFs do not match up with those of the PDF. 
Alternatively, if these are not considered IR divergences and absorbed into the BF's renormalization factor  (as in a pure dimensional regularization calculation) the matching onto PDFs is still feasible. In this approach there is a lack of distinction between rapidity and UV divergences.
However, this alternative does not apply to our BFs, since there is a definite measurement that restricts the rapidity and prohibits rapidity divergences.

In order to perform the matching, we need also the PDFs calculated with $m$ and $\de$ regulators. Our final result is
\begin{align}
f_{q/ q} &= \frac{\alpha _s(\mu)\,C_F}{2\pi }\theta(x) \bigg\{\Big(\frac{1}{\epsilon }+\ln \frac{\mu ^2}{m^2}-\ln x \Big)\Big[2x \cL_0(1-x) +\frac{3}{2}\delta (1-x) + \theta(1-x)(1-x) \Big] 
\nn \\ & \hspace{2 cm} 
+\Big(\frac{9}{4}-\frac{\pi ^2}{3}\Big)\delta (1-x) -2\theta(1-x)(1-x) \bigg\}  
\,, \nn \\
f_{q/ g} &=\frac{\alpha _s(\mu)\,T_F}{\pi }\theta(x) \bigg[ P_{qg}(x) \bigg( \frac{1}{\epsilon }+\ln \frac{\mu ^2}{m^2} \bigg ) -\theta(1-x)\bigg] \, ,
\end{align}
where the $\de$-dependence cancels. Here $m$ denotes the gluon mass in $f_{q/q}$ and the quark mass in $f_{q/g}$ as was the case for beam function. We obtain the same matching coefficients as in \sec{oneloop}.

\section{One-Loop Gluon BF in Momentum Space}
\label{app:gfupdf}

\begin{figure}[t]
\centering
\includegraphics[width=\textwidth]{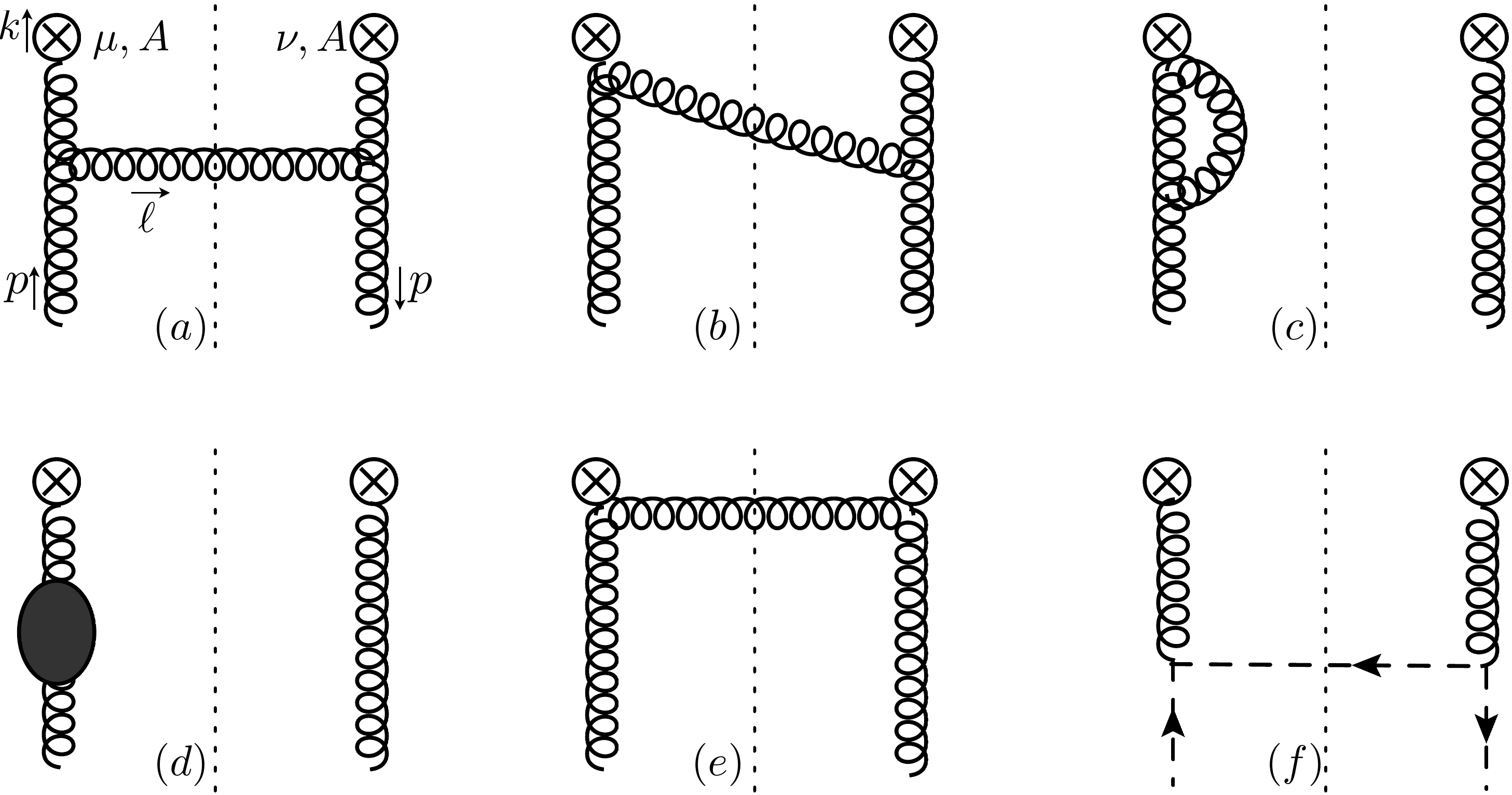}
\caption{Feynman diagrams contributing to gluon BF at one-loop. Diagrams (b) and (c) have symmetric counterparts that are not shown. Diagram (d) represents self-energy corrections. Diagram (b) vanishes because we restrict the polarization sum to physical polarizations. Diagrams (c) and (d) vanish in pure dimensional regularization and diagram (e) vanishes in the Feynman gauge.
}
\label{fig:gfupdf}
\end{figure}

In this section we calculate the gluon BF with uncontracted indices $\mu,\nu$, using CDR${}_2$. The diagrams are shown in \fig{gfupdf}. We work in pure dimensional regularization, which sets the virtual diagram (c) and the wave function diagram (d) to zero because they contain no dimensionful quantity. We will use Feynman gauge, which makes diagram (e) vanish because the gluons emitted by the Wilson lines have $\bn$-polarization and $\bn \cdot \bn = 0$. By only summing over the physical polarizations of the final state gluon [see \eq{polsum}], diagram (b) vanishes as well. Thus only diagram (a) contributes to $B_g^g$ and only (f) contributes to $B_g^q$.

We will start with diagram \fig{gfupdf}(a), first giving the expressions and then we will comment on the calculation and several of its subtleties.
\begin{align}
B_{g,\bare}^{g(a)} & = - \Big(\frac{e^{\ga_E} \mu^2}{4\pi}\Big)^\eps \theta(x)\! \int\! \frac{\df^d \ell}{(2\pi)^d}\, 2\pi \theta(\ell^0) \de(\ell^2) \, \frac{\de^{AB}}{N_c^2-1}
\frac{1}{d-2} \sum_\text{pol} \ve_\al(p) \ve_\bt^*(p)\, \sum_\text{pol} \ve_\la^*(\ell) \ve_\rho(\ell) 
\nn \\ & \quad \times
g f^{AEC} \big[g^{\al \la} (\ell + p)^\ga - g^{\ga \la} (2\ell-p)^\al + g^{\al \ga} (\ell-2p)^\la\big] \,
\nn \\ & \quad \times
g f^{BDE} \big[ g^{\bt\rho}(\ell+p)^\de - g^{\de \rho}(2\ell-p)^\bt + g^{\bt \de}(\ell-2p)^\rho\big]
\nn \\ & \quad \times
\de^{CD} \Big[g_\perp^{\mu\ga} \!-\! \frac{\ell_\perp^\mu \bn^\ga}{\bn \sdt (\ell\!-\!p)}\Big] 
\Big[g_\perp^{\nu\de} \!-\! \frac{\ell_\perp^\nu \bn^\de}{\bn \sdt (\ell\!-\!p)}\Big] \Big[\frac{-\img}{(p\!-\!\ell)^2}\Big]^2 
\de(p^- \!\!-\! \ell^- \!\!-\! k^-) \de(\ell^+ \!+\! k^+) \de^2(\vec \ell_\perp \!+\! \vec k_\perp)
\nn \\ & 
= \frac{\al_s C_A}{(d-2)\pi^2}\, \theta(x)\theta(1-x)\, \Big(\frac{e^{\ga_E} \mu^2}{4\pi}\Big)^\eps \int\! \frac{\df^{-2\eps} \ell_\eps}{(2\pi)^{-2\eps}}\, 
\de\Big( \frac{1-x}{x} t - \vks + \ell_\eps^2 \Big) \frac{1}{t}
\nn \\ & \quad \times
\bigg\{ \Big[x(1-x) + \frac{x}{1-x} \Big] g_\perp^{\mu\nu}
-\frac{d-2}{t} (k_\perp^\mu k_\perp^\nu + \ell_\eps^\mu \ell_\eps^\nu)\bigg\}
\nn \\ & 
= \frac{\al_s C_A}{(d - 2)\pi^2}\, \theta(x)\, \frac{e^{\eps \ga_E} \mu^{2\eps}}{\Ga(-\eps)} \theta(t)t^{-1-\eps}\, \theta(1-u)(1-u)^{-1-\eps}\,
\theta(1-x)\Big(\frac{x}{1-x}\Big)^\eps 
\nn \\ & \quad \times
\bigg\{\Big[x(1-x) + \frac{x}{1-x} \Big] g_\perp^{\mu\nu}
- (d-2) u \frac{1-x}{x}\, \frac{k_\perp^\mu k_\perp^\nu}{\vks} - \frac{1-\eps}{\eps} (1-u) \frac{1-x}{x} g_\eps^{\mu\nu} \bigg\} \frac{\df u}{\df \vks}
\nn \\ &
= \frac{\al_s C_A}{2\pi}\,\theta(x) \bigg( \bigg\{\frac{2}{\eps^2} \de(t) \de(1\!-\!x) \!-\! \frac{1}{\eps} \Big[ \frac{2}{\mu^2} \cL_0\Big(\frac{t}{\mu^2}\Big) \de(1\!-\!x) \!+\! \de(t) P_{gg}(x) \Big] 
\!+\! \frac{2}{\mu^2} \cL_1\Big(\frac{t}{\mu^2}\Big) \de(1\!-\!x) 
\nn \\ & \quad
+ \frac{1}{\mu^2} \cL_0\Big(\frac{t}{\mu^2}\Big) P_{gg}(x)
+ \de(t) \Big[ \cL_{1}(1\!-\!x)\frac{2(1\!-\!x\!+\!x^2)^2}{x} 
\!-\! \ln x \, P_{gg}(x) \!-\! \frac{\pi^2}{6} \de(1\!-\!x) \Big]
\bigg\} \frac{g_\perp^{\mu\nu}}{d-2}
\nn \\ & \quad 
- \frac{2}{\mu^2} \cL_0\Big(\frac{t}{\mu^2}\Big)\, \theta(1-x) \frac{1-x}{x} \bigg[\frac{k_\perp^\mu k_\perp^\nu}{\vks} + \frac{g_2^{\mu\nu}}{2} \bigg]
\bigg)\,
\frac{1}{\pi} \de\Big( \vks - \frac{1-x}{x} t \Big)
\,.\end{align}
The plus distributions $\cL_{n}$ are defined in \eq{plusdef} and the splitting function $P_{gg}$ in \eq{split}. Following the standard gluon BF calculation in ref.~\cite{Berger:2010xi}, we will average over the colors and polarizations of the incoming gluons,
\begin{align}
  \frac{1}{d-2} \sum_\text{pol} \ve_\al^*(p) \ve_\bt(p) = -\frac{g^\perp_{\al\bt}}{d-2}
  \,. \quad  
\end{align}
For the intermediate gluons we restrict the polarization sum to physical polarizations,
\begin{align} \label{eq:polsum}
  \sum_\text{pol} \ve^*_\la(\ell) \ve_\rho(\ell) &= 
  -g^\perp_{\la\rho} + \frac{\bn_\la \ell^\perp_\rho}{\bn \sdt \ell} + \frac{\ell^\perp_\la \bn_\rho}{\bn \sdt \ell} - \frac{\bn_\la \bn_\rho \ell_\perp^2}{(\bn \sdt \ell)^2}
\,.
\end{align}
The polarization sum for the incoming gluons is simpler because $p_\perp^\mu=0$. 

In the first step we performed all the Lorentz contractions and integrated most of the delta functions. It is convenient to decompose $\ell_\perp^\mu = \ell_2^\mu + \ell_\eps^\mu$, where $\ell_2$ lives in 2 dimensions and $\ell_\eps$ in $-2\eps$ dimensions. In CDR${}_2$, the transverse momentum measurement does not constrain
$\ell_\eps$ and its integral produces the required divergences,
\begin{equation}
  \int\! \frac{\df^{-2\eps} \ell_\eps}{(2\pi)^{-2\eps}}
  = \frac{(4\pi)^\eps}{\Ga(-\eps)} \int\! \df(-\ell_\eps^2)\, (-\ell_\eps^2)^{-1-\eps}
\,.\end{equation}
The $\ell_\eps^\mu \ell_\eps^\nu$ Lorentz structure produces a $-\ell_\eps^2\, g_\eps^{\mu \nu}/(2\eps)$, where $g_\eps^{\mu\nu}$ is the $-2\eps$-dimensional part of the metric tensor.

To clearly separate the variables in the third expression, we swapped $\vks$ for $u = x\, \vks/[(1-x)\, t]$, which is kinematically restricted to $0 \leq u \leq 1$. We included the Jacobian for changing the variable of the distribution from $\vks$ to $u$, since distributions are tied to the corresponding measures,
\begin{align}
  B(\vks,\dots) \df \vec k_\perp^2= B(u,\dots) \df u\,.
\end{align}
We then use the distribution identity
\begin{equation} \label{eq:distr_id}
 \frac{\theta(z)}{z^{1+\eps}} = - \frac{1}{\eps}\,\delta(z) + \cL_0(z)
  - \eps \cL_1(z) + \ord{\eps^2}
\,,\end{equation}
to obtain distributions in $t/\mu^2$, $1-x$ and $1-u$. (The $\df u/\df \vks$ should not be expanded.) Based on the one-loop kinematics we would expect the expression to be proportional to $\de(1-u)$, but the expansion also produces a $\cL_0(1-u)$ and $\cL_1(1-u)$ as well. However, all these terms are multiplied by a $\de(t)$ or a $\de(1-x)$, implying that $\vks = 0$. The situation is similar to having a $\phi$ dependence at $\theta=0$ in spherical coordinates, where all choices of $\phi$ are equivalent. 
We are therefore free to replace this $u$ dependence by a delta function times its average value
\begin{equation}
  \cL_n(1-u) \to 0
  \,, \qquad
  u \cL_n(1-u) \to \de(1-u)\, \int_0^1\! \df v\, v \cL_n(1-v) = (-1)^{n+1} n!\, \de(1-u)
\,.
\end{equation}
After this, we change variables back from $u$ to $\vks$, where the Jacobian $\df u/\df \vks$ gets absorbed into $\de(1-u)$ to yield $\de[\vks-(1-x)t/x]$ in the final expression.

We write the final result in terms of independent Lorentz structures. For the new tensor term we have
\begin{equation}
  \de(\vks)\, \Big(\frac{k_\perp^\mu k_\perp^\nu}{\vks} + \frac{g_2^{\mu\nu}}{2} \Big)
  = 0
  \,,
\end{equation}
because if $\vks = 0$ there is no preferred direction and $k_\perp^\mu k_\perp^\nu/\vks \to -g_2^{\mu\nu}/2 $. This observation eliminates the divergent contribution for this Lorentz structure, which cannot be present because this tensor first appears at one loop. In the final result the expressions multiplying $g_\perp^{\mu\nu}$ are the same as for the standard BF \cite{Stewart:2010qs,Berger:2010xi}. The $g_\eps^{\mu\nu}$ structure, that appeared in intermediate expressions, is absorbed into this piece. This is not surprising because there is no preferred direction in the $-2\eps$-dimensional part of space. For all intents and purposes, we could therefore have taken the $\mu$ and $\nu$ indices in 2 dimensions from the beginning [the factor of $1/(1-\eps)$ in $g_\perp^{\mu\nu}/(1-\eps)$ will never affect the renormalization nor the matching]. We reproduce the known renormalization and  matching for the $g_\perp^{\mu\nu}$ piece and obtain the matching coefficient for the $k_\perp$-dependent Lorentz structure.

The calculation of the diagram in \fig{gfupdf}(f) with external quarks instead of gluons is very similar
\begin{align}
B_{g,\bare}^q & = - \Big(\frac{e^{\ga_E} \mu^2}{4\pi}\Big)^\eps \theta(x)\! \int\! \frac{\df^d \ell}{(2\pi)^d}\, 2\pi \theta(\ell^0) \de(\ell^2) \, 
\frac{1}{2}  \tr \Big[ \sum_\text{spins} u(p) \bar u(p)\, \img gT^D \Big(n_\de + \frac{\ga_{\perp\de} \ellslash_\perp}{\bn \sdt \ell}\Big) \frac{\bnslash}{2} 
\nn \\ & \quad \times
\sum_\text{spins} u(\ell) \bar u(\ell) \img gT^C \Big(n_\ga + \frac{\ellslash_\perp \ga_{\perp\ga}}{\bn \sdt \ell}\Big) \frac{\bnslash}{2} \Big]\,
\de^{CD} \Big[g_\perp^{\mu\ga} - \frac{\ell_\perp^\mu \bn^\ga}{\bn \sdt (\ell-p)}\Big] 
\Big[g_\perp^{\nu\de} - \frac{\ell_\perp^\nu \bn^\de}{\bn \sdt (\ell-p)}\Big] 
\nn \\ & \quad \times
\Big[\frac{-\img}{(\ell-p)^2}\Big]^2\,
\de(p^- \!- \ell^- \!- k^-) \de(\ell^+ + k^+) \de^2(\ell_\perp + k_\perp)
\nn \\ &
= \frac{\al_s C_F}{4\pi^2}\, \theta(x)\, \frac{e^{\eps \ga_E}}{\Ga(-\eps)}\, \theta(t)t^{-1-\eps}\, \theta(1-u)(1-u)^{-1-\eps}\,
\nn \\ & \quad \times
\theta(1-x)\Big(\frac{x}{1-x}\Big)^\eps 
\bigg[x g_\perp^{\mu\nu}
- 4u\, \frac{1-x}{x} \frac{k_\perp^\mu k_\perp^\nu}{\vks} - \frac{2}{\eps}(1-u) \frac{1-x}{x} g_\eps^{\mu\nu} \bigg] \frac{\df u}{\df \vks}
\nn \\ &
= \frac{\al_s C_F}{2\pi}\, \theta(x)\,\bigg(\bigg\{ \Big[-\frac{1}{\eps} \de(t) +\frac{1}{\mu^2} \cL_0\Big(\frac{t}{\mu^2}\Big) +
\de(t) \ln \frac{1\!-\!x}{x}\Big] P_{gq}(x) + \de(t)\, \theta(1\!-\!x)x \bigg\} \frac{g_\perp^{\mu\nu}}{d-2}
\nn \\ & \quad
- \frac{2}{\mu^2} \cL_0\Big(\frac{t}{\mu^2}\Big)\, \theta(1-x)\frac{1-x}{x} \bigg[\frac{k_\perp^\mu k_\perp^\nu}{\vks} + \frac{g_2^{\mu\nu}}{2} \bigg] \bigg)\, \frac{1}{\pi}\de\Big( \vks - \frac{1-x}{x} t \Big)
\,.\end{align}
This result also agrees with the matching obtained from the standard beam function calculation. All the matching coefficients are collected in \sec{gBF-matching}.

\bibliographystyle{jhep}
\bibliography{../Fragmentation}


\end{document}